\documentclass[aps, 12pt, preprint, preprintnumbers, amsmath,amssymb,nofootinbib,superscriptaddress,hyperref]{revtex4-1} 
\usepackage{amssymb}
\usepackage{slashed}
\usepackage{amsmath}
\usepackage{braket}
\usepackage{float}
\usepackage[utf8]{inputenc}
\usepackage{graphicx}
\usepackage{subfigure}
\usepackage{latexsym}
\usepackage{epic}
\usepackage{eepic}
\usepackage{epsfig}
\usepackage{color}
\usepackage{natbib}
\usepackage[colorlinks=true,citecolor=darkred,urlcolor=darkred, pdfborder={0 0 0}]{hyperref}
\usepackage[normalem]{ulem}

\definecolor{darkred}{rgb}{0.6,0,0}

\newcommand{\bea}{\begin{eqnarray}}
\newcommand{\eea}{\end{eqnarray}}

\newcommand{\AddrAHEP}{
  AHEP Group, Institut de F\'{i}sica Corpuscular,
  CSIC/Universitat de Val\`{e}ncia, Parc Cient\'ific de Paterna.\\
 C/ Catedr\'atico Jos\'e Beltr\'an, 2 E-46980 Paterna (Valencia), Spain}

\definecolor{linkcolor}{rgb}{0,0,0.5}

\begin{document}
\title{Testing triplet fermions at the electron-positron and electron-proton colliders using fat jet signatures}

\author{Arindam Das}\email{arindam.das@het.phys.sci.osaka-u.ac.jp}
\affiliation{Department of Physics, Osaka University, Toyonaka, Osaka 560-0043, Japan}
\author{Sanjoy Mandal}\email{smandal@ific.uv.es}
\affiliation{\AddrAHEP}
\author{Tanmoy Modak}
\email{tanmoyy@hep1.phys.ntu.edu.tw}
\affiliation{Department of Physics, National Taiwan University, Taipei 10617, Taiwan}
\preprint{\textbf{OU-HET-1055}}
 %%%%%%%%%%%%%%%%%%%%%%%%%%%%%%%%%%%%%%%%
\bibliographystyle{unsrt} 
%%%%%%%%%%%%%%%%%%%%%%%%%%%%%%%%%%%%%%%%%%%%%
\begin{abstract}
The addition of $SU(2)_L$ triplet fermions of zero hypercharge with the Standard Model (SM) helps to explain the origin of the neutrino mass by the so-called seesaw mechanism. Such a scenario is commonly know as the type-III seesaw model. 
After the electroweak symmetry breaking the mixings between the light and heavy mass eigenstates of the neutral leptons are developed 
which play important roles in the study of the charged and neutral multiplets of the triplet fermions at the colliders.  
In this article we study such interactions to produce these multiplets of the triplet fermion at the electron-positron and electron-proton colliders at different center of mass energies.
We focus on the heavy triplets, for example, having mass in the TeV scale 
so that their decay products including the SM the gauge bosons or Higgs boson can be sufficiently boosted, leading to a fat jet.
Hence we probe the mixing between light-heavy mass eigenstates of the neutrinos and compare the results 
with the bounds obtained by the electroweak precision study. 
\end{abstract}
\vspace{-3cm}
\maketitle
%\tableofcontents
%%%%%%%%%%%%%%%%%%%%%%%%%%%%%%%%%%%%
\section{Introduction}
\label{introduction}
%%%%%%%%%%%%%%%%%%%%%%%%%%%%%%%%
The observations of the neutrino oscillation phenomena and the flavor mixing \cite{Anselmann:1995ag,Fukuda:1996sz,Athanassopoulos:1997pv, Fukuda:1998tw,Abdurashitov:1994bc,Ahmad:2002jz,Ahmad:2002ka,Hosaka:2006zd,Eguchi:2002dm,Ahn:2002up,Catanesi:2013fxa,Adamson:2011qu,Abe:2011fz,An:2012eh,Ahn:2012nd,Patrignani:2016xqp} give a very strong indication of the existence of the tiny neutrino masses. The existence of the neutrino mass is one of the missing pieces in the SM. Therefore to correctly fit the experimental results in, the SM requires to be extended. From the point of view of the low energy theory, one can introduce a dimension-$5$ operator within the SM which involves the SM Higgs and lepton doublets. Such an operator is known as the very famous Weinberg operator \cite{Weinberg:1979sa} which introduces a violation of the lepton number by two units. The breaking of the electroweak symmetry ensures the generation of the tiny neutrino Majorana masses which is suppressed by the scale of the dimension-$5$ operator. In the context of the renormalizable theory, the dimension-$5$ operator is naturally generated by the inclusion of the SM singlet right handed Majorana neutrinos and finally integrating them out. This is called the seesaw mechanism \cite{Minkowski:1977sc,Mohapatra:1979ia,Schechter:1980gr,Yanagida:1979as,GellMann:1980vs,Glashow:1979nm} which can successfully explain the origin of the tiny neutrino mass, however, hitherto there is no experimental evidence of this simple but extraordinary theoretical aspect. Such non observation opens up the pathway for a variety of neutrino mass generation mechanisms. 

Type-III seesaw scenario is one of the most simple scenarios where the SM is extended by an SU$(2)_L$ right handed triplet fermion with zero hypercharge \cite{Foot:1988aq} which effectively generates a lepton number violating dimension-$5$ Weinberg operator. The triplet fermion has neutral and charged multiplets. The neutral component gets involved in the generation of the Majorana mass term for the light neutrinos after the electroweak symmetry breaking which finally generates the mixing between the light-heavy mass eigenstates  as it happens in the canonical or type-I seesaw mechanism. These neutral multiplets can be studied at the energy frontier form their production at the different colliders through the mixing. Apart from the neutral multiplets the charged multiplets can also be produced at the collider in the same fashion, however, the charged multiplets can also be produced directly in pair from the SM gauge interactions where a wide variety of phenomenological aspects can be studied involving the prompt and non-prompt decay of the triplet fermions \cite{delAguila:2008hw,delAguila:2008cj,Franceschini:2008pz,Biggio:2011ja,Bandyopadhyay:2011aa,Ruiz:2015zca,Goswami:2017jqs,Jana:2019tdm}. 

The rich theoretical and phenomenological aspects of the type-III seesaw scenario has been explored in different ways. An SU$(5)$ realization of this scenario from the $\textbf{24}$ representation containing both of a triplet fermion and a singlet fermion has been proposed in \cite{Ma:1998dn}. A supersymmetric realization of the SU$(5)$ grand unified theories for the singlet and triplet has been studied in \cite{Dimopoulos:1981zb,Sakai:1981gr}. In this case the triplet can reside in the intermediate scale to reproduce the neutrino oscillation data. A nonsupersymmetric implementation of the type-III seesaw in the framework of SU$(5)$ grand unified theory with the inclusion of the adjoint fermionic multiplet has been studied in \cite{Bajc:2006ia} which mainly predicts a theory of the light SU$(2)$ triplet fremion with mass at the electroweak scale. It has been mentioned in \cite{Bajc:2006ia} that due to the gauge coupling of the multiplets of the triplets, they can be pair-produced directly through the Drell-Yan process. Being Majorana in nature, the neutral component of the triplets can show up with a distinct lepton number violating signature at the collider. The grand unified theory inspired non-supersymmetric and supersymmetric renormalizable SU$(5)$ frameworks to study the origin of neutrino masses generated by type-III and type-I seesaw mechanisms have been discussed in \cite{Perez:2007rm, Perez:2007iw}. Collider phenomenology of the heavy triplet fermions from such models have been studied in \cite{Adhikari:2008uc,Arhrib:2009mz,Bajc:2009ft}. An inverse seesaw realization under the type-III seesaw framework has been studied in \cite{Ibanez:2009du} involving a singlet hyperchargeless fermion. Another type of type-III seesaw was studied in \cite{Ma:2009kh} where an extra U$(1)$ gauge group has been introduced to the SM under the anomaly free scenario \cite{Barr:1986hj,Ma:2001kg,Barr:2005je}.

The renormalization group evolution of the effective neutrino mass matrix in the SU$(2)_L$ triplet fermion extended SM with emphasis on the threshold effects has been studied in \cite{Chakrabortty:2008zh}. In the type-III seesaw  scenario with degenerate heavy triplets the impact of the renormalization group evolution in the context of perturbativity bounds and vacuum stability of the scalar potential has been studied in \cite{Gogoladze:2008ak}. Adding two SU$(2)_L$ triplet fermions with zero hypercharge has been studied to probe electroweak vacuum stability with the nonzero neutrino mass, naturalness and lepton flavor violation in \cite{Goswami:2018jar}. A simple realization of the triplet fermion under SU$(2)_L$ with a general U$(1)$ extension of the SM has been discussed in \cite{Ma:2002pf} where a neutral gauge boson $(Z^\prime)$ plays an important role for the triplet fermion in addition to the SM gauge boson mediated processes. Over the years at different center of mass energies and integrated luminosities, LHC is also searching for the fermions triplet under the SU$(2)_L$ group in \cite{CMS:2012dza,CMS:2012ra,CMS:2015mza,CMS:2016hmk,Sirunyan:2017qkz,CMS:2017wua,CMS:2019xud,Aad:2015cxa,ATLAS:2018ghc} from SM gauge mediated processes using different flavor structures of the Yukawa interaction involving the triplet fermion, SM lepton and Higgs doublets. The latest result from the LHC can be found in  which put strong bounds on the triplets lighter than $880$ GeV \cite{CMS:2019xud} and $590$ GeV\cite{ATLAS:2018ghc}. This leads us to find a strategy for the heavy triplet fermion search at the energy frontier in the future.

Apart from the collider searches, Lepton Flavor Violating (LFV) \cite{Abada:2008ea} and non-unitarity effects in the type-III model have been studied in \cite{Abada:2007ux,Biggio:2019eeo}. In this context it is necessary to mention that such studies have been performed in the context of the type-I seesaw scenario \cite{Antusch:2007zza,Forero:2011pc,Basso:2013jka,Antusch:2014woa,Fernandez-Martinez:2015hxa,Antusch:2016brq,Fong:2017gke,Hernandez-Garcia:2017pwx,Escrihuela:2019mot}. The bounds on the light heavy neutrino mixings from the Eletroweak Precision Data (EWPD) have been studied in \cite{Raidal:2008jk,delAguila:2008pw} which can be considered as the upper limits to constrain the limits on the mass-mixing plane for the triplets.

In this paper we study the production of the TeV scale triplet fermions at electron-positron $(e^-e^+)$ collider. At the $e^-e^+$ collider the production of the neutral multiplet of the triplet fermion will take place through the $s$ and $t$ channel process in association with the SM leptons. Similarly the charged multiplets of the triplet can be produced from the $s$ and $t$ channel processes in association with the SM leptons, however, they can be produced in pair from the $s$ channel process. If the triplets are in the TeV scale and heavy they can sufficiently boost the decay products such as SM gauge bosons $(W, Z)$ and  the SM Higgs $(h)$ which can further produce fat jets from the leading hadronic decay modes. Such a fat jet signature can have a distinctive nature to separate the signal from the SM backgrounds. Following this we study the allowed parameter space in the mass-mixing plane from the associated production of the triplets with SM leptons. We also study the significance of several final states coming from the pair production of the charged multiplets. At the $e^+ e^-$ collider we consider two center of mass energies $\sqrt{s}=1$ TeV and $3$ TeV with integrated luminosities at $\mathcal{L}=1$ ab$^{-1}$, $3$ ab$^{-1}$ and $5$ ab$^{-1}$, respectively. In this context we mention that in the seesaw scenario a variety of final states have been studied for the linear collider in the literatures \cite{Gluza:1995ix,Gluza:1995js,Gluza:1997ts,Das:2012ze,Banerjee:2015gca,Antusch:2015gjw,Antusch:2016vyf,Antusch:2016ejd,Das:2017nvm,Chakraborty:2018khw,Das:2018usr}. We also study the production of the neutral and charged multiplet of the triplet fermion at the electron-proton $(e^-p)$ collider at different center of mass energies, $\sqrt{s}=1.3$ TeV, $1.8$ TeV and $3.46$ TeV. As we are concentrating in the heavy mass range of the triplet fermions we expect that their decay products will be sufficiently boosted to probe the mass mixing plane. To do this we fix the collider energy at $\sqrt{s}=3.46$ TeV (FCC-he) with the luminosity $\mathcal{L}=1$ ab$^{-1}$, $3$ ab$^{-1}$ and $5$ ab$^{-1}$, respectively. Studies on the $e^-p$ collider considering the seesaw scenario have been performed in \cite{Mondal:2015zba,Mondal:2016kof,Mandal:2018qpg}. In both of these colliders to identify the fat jet from the boosted decay products of the heavy triplet fermions hence we  study the signals and the SM backgrounds. In this context we mention that in the $e^-p$ collider testing the Majorana nature of a fermion could be very interesting because it can show a lepton number violation signature distinctively. Such a process for the  seesaw scenario has been studied in \cite{Antusch:2016ejd,Das:2018usr}.  

This paper is organized as follows. In Sec.~\ref{model}, we discuss the model and the interactions of the triplet fermions with the SM particles. 
In Sec.~\ref{production} we study various production processes of the charged and neutral multiplets of the triplet fermion at the $e^-e^+$ and $e^-p$ colliders. 
In Sec.~\ref{analysis} we discuss the complete collider study of various possible final states and calculate the bounds on the mixing angles in the mass-mixing plane. 
We also compare the results with the existing constraints. In Sec.~\ref{Dis} we discuss the calculated bounds.
Finally, in Sec.~\ref{conclusion} we conclude.
%%%%%%%%%%%%%%%%%%%%
\section{Model}
\label{model}
%%%%%%%%%%%%%%%%%%%%%%%%%%%%%%%%%%%%%%
The type-III seesaw model is a simple extension of the SM with an $SU(2)_L$ triplet fermion $(\tilde{\Sigma})$ with zero hypercharge. 
A three generation triplet fermion of this kind can be introduced in this model to generate the neutrino mass. 
For simplicity we suppress the generation indices. The relevant part of the Lagrangian can be written as 
\begin{align}
\mathcal{L}_{\text{int}}=\mathcal{L}_{\text{SM}}+\text{Tr}(\overline{\tilde{\Sigma}}i \gamma^\mu D_\mu \tilde{\Sigma})-\frac{1}{2}M_{\Sigma}\text{Tr}(\overline{\tilde{\Sigma}}\tilde{\Sigma}^c+\overline{\tilde{\Sigma}^c}\tilde{\Sigma})-\sqrt{2}(\overline{\ell_L}Y^\dagger \tilde{\Sigma} H + H^\dagger \overline{\tilde{\Sigma}} Y \ell_L)
\label{L}
\end{align}
where the first term is the kinetic interaction of the triplet and $D_\mu$ represents the covariant derivative. $\mathcal{L}_{\text{SM}}$ is relevant part of the SM Lagrangian. In the second term of Eq.~\ref{L}, $M_{\Sigma}$ represents the triplet mass parameter. For simplicity we consider $M_\Sigma$ as a real parameter and the triplets are degenerate in nature. Therefore $M_\Sigma$ is a real diagonal matrix. $Y$ in the third term of Eq.~\ref{L} is the Yukawa coupling between the SM lepton doublet $(\ell_L)$, SM Higgs doublet $(H)$ and the triplet fermion $(\tilde{\Sigma})$. In this analysis we represent the the relevant SM candidates in the following way 
\bea
\ell_L
 =
 \begin{pmatrix}
  \nu_{L}\\
  e_{L} \\
 \end{pmatrix}\,\,\,\text{and}\,\,\,
H = 
 \begin{pmatrix}
 \phi^0\\
  \phi^-\\
 \end{pmatrix}.
 \label{L2}
\eea
After the symmetry breaking $\phi^0$ acquires the vacuum expectation value and we can express it as $\phi^0=\frac{v+h}{\sqrt{2}}$ with $v=246$~GeV.
The explicit form of $\tilde{\Sigma}$ and its charge conjugate $\tilde{\Sigma}^c\equiv C\overline{\tilde{\Sigma}}^T$ ($C$ is the charge conjugation operator) are given by
\begin{align}
 \tilde{\Sigma}=
 \begin{pmatrix}
  \Sigma^0/\sqrt{2}  &  \Sigma^+ \\
\Sigma^-           &   -\Sigma^0/\sqrt{2}  \\
 \end{pmatrix}\,\,\,\text{and}\,\,\,\,
  \tilde{\Sigma}^c=
 \begin{pmatrix}
  \Sigma^{0c}/\sqrt{2}  &  \Sigma^{-c} \\
  \Sigma^{+c}           &   -\Sigma^{0c}/\sqrt{2}  \\
 \end{pmatrix}
\end{align}
whereas $D_\mu$ is given by
\begin{align}
 D_\mu=\partial_\mu-i\sqrt{2}g
 \begin{pmatrix}
  \frac{W^3_\mu}{\sqrt{2}} & W_{\mu}^+ \\
  W_\mu^- &  -\frac{W_\mu^3}{\sqrt{2}} \\
 \end{pmatrix}
\end{align}
To study the mixing between the SM charged leptons and the charged multiplets of the triplet fermions, it is 
convenient to express the four degrees of freedom of each of the charged multiplets of the triplet fermions in terms of a single Dirac
spinor as $\Sigma=\Sigma_R^-+\Sigma_R^{+c}$. The neutral multiplets of the triplet fermions are two component fermions. 
Therefore they have the two degrees of freedom. Finally mixing with the SM light neutrinos, these neutral fermions generate tiny neutrino mass through the seesaw mechanism after the electroweak symmetry breaking. With this convention we re-write the Eq.~\ref{L} as
\bea
\mathcal{L}_{\text{int}}&=&\overline{\Sigma}i\slashed{\partial}\Sigma+\overline{\Sigma^0_R} i \slashed{\partial}\Sigma^0_R-\overline{\Sigma}M_{\Sigma}\Sigma-\left(\overline{\Sigma^0_R}\frac{M_\Sigma}{2}\Sigma_R^{0c}+\text{H.c.}\right) \nonumber \\ 
 &-&\bigg(\phi^0\overline{\Sigma_R^0} Y_\Sigma\nu_L+ \sqrt{2}\phi^0\overline{\Sigma}Y e_L+\phi^+ \overline{\Sigma_R^0} Y e_L-\sqrt{2}\phi^+\overline{\nu_L^c} Y^T\Sigma+\text{H.c.}\bigg) \nonumber \\
 &-&g W_\mu^3\overline{\Sigma}\gamma^{\mu}\Sigma
 \label{Lm}
\eea
After the electroweak symmetry breaking we can derive the mass matrix for charged and neutral sectors.
The mass term of the charged leptons shows the usual nature of the Dirac particles and it can be written as  
\begin{align}
 -\mathcal{L}_{\text{mass}}^{\text{charged}}=
 \begin{pmatrix}
  \overline{e_R} & \overline{\Sigma}_R \\
 \end{pmatrix}
 \begin{pmatrix}
  m_\ell & 0 \\
  Y v & M_\Sigma \\
 \end{pmatrix}
 \begin{pmatrix}
  e_L \\
  \Sigma_L \\
 \end{pmatrix}
 +
 \begin{pmatrix}
  \overline{e}_L & \overline{\Sigma}_L \\
 \end{pmatrix}
 \begin{pmatrix}
  m_\ell & Y^\dagger v\\
  0 & M_\Sigma \\
 \end{pmatrix}
 \begin{pmatrix}
  e_R \\
  \Sigma_R \\
 \end{pmatrix} 
 \label{ml}
\end{align}
%%%%%%%%%%%%%%%%%
where $m_\ell$ represents the Dirac mass of the SM charged lepton coming from the well known SM Lagrangian. 
Similarly the mass term for the neutral fermions can be given as 
%%%%%%%%%%%%%%%%%%
\begin{align}
 -\mathcal{L}^{\text{neutral}}_{\text{mass}}&=
 \begin{pmatrix}
  \overline{\nu_L} & \overline{\Sigma_R^{0c}} \\
 \end{pmatrix}
 \begin{pmatrix}
  0 & Y^{\dagger}\frac{v}{2\sqrt{2}} \\
  Y^{*}\frac{v}{2\sqrt{2}} & \frac{M_\Sigma}{2} \\
 \end{pmatrix}
 \begin{pmatrix}
  \nu_L^c \\
  \Sigma_R^0 \\
 \end{pmatrix}
 +
 \begin{pmatrix}
  \overline{\nu_L^c} & \overline{\Sigma_R^0} \\
 \end{pmatrix}
 \begin{pmatrix}
  0 & Y^T \frac{v}{2\sqrt{2}} \\
  Y\frac{v}{2\sqrt{2}} & \frac{M_\Sigma}{2} \\
 \end{pmatrix}
 \begin{pmatrix}
  \nu_L \\
  \Sigma_R^{0c} \\
 \end{pmatrix}
 \label{n1}
\end{align}
%%%%%%%%%%%%%%%%%%%%%%%%%%%%%%%%%%%%%%
In case of a Dirac mass, the charged lepton mass matrix can be diagonalized by a bi-unitary transformation which can be written as
%%%%%%%%%%%%%%%
 \bea
 \begin{pmatrix}
  e_{L} \\
  \Sigma_{L} \\
 \end{pmatrix}
 =\mathcal{U}_{L}
 \begin{pmatrix}
  e^\prime_{L}\\
  \Sigma^\prime_{L} \\
 \end{pmatrix} \, \, \, \, \text{and} \, \, \, \,
 \begin{pmatrix}
  e_{R} \\
  \Sigma_{R} \\
 \end{pmatrix}
 =\mathcal{U}_{R}
 \begin{pmatrix}
  e^\prime_{R}\\
  \Sigma^\prime_{R} \\
 \end{pmatrix}. 
 \label{U1}
\eea
The transformation matrices in Eq.~\ref{U1} can be written as 
\bea
\mathcal{U}_L = 
 \begin{pmatrix}
  1-\epsilon& Y^\dagger M_\Sigma^{-1} v\\
  -M_\Sigma^{-1} Y v& 1-\epsilon^{\prime} 
 \end{pmatrix} \,\,\,\, \text{and} \,\,\,
\mathcal{U}_R = 
 \begin{pmatrix}
  1& m_\ell Y^\dagger M_\Sigma^{-2} v\\
 -M_\Sigma^{-2} Y m_\ell v& 1
 \end{pmatrix} 
 \label{U11}
  \eea
%%%%%%%%%%%%%%%%%%%%%%%%%%
The symmetric neutral lepton mass matrix can be diagonalized by a single unitary which can be written as 
\bea
\begin{pmatrix}
  \nu_L \\
  \Sigma_R^{0c} \\
 \end{pmatrix}
 =\mathcal{V}
 \begin{pmatrix}
  \nu^\prime_L\\
  \Sigma_R^{\prime 0c}\\
 \end{pmatrix}
 \label{U2}
\eea
%%%%%%%%
The transformation matrix in Eq.~\ref{U2} can be written as 
%%%%%%%%%%%
\bea
\mathcal{V} = 
 \begin{pmatrix}
 (1-\frac{\epsilon}{2}) U_{\text{PMNS}}& \frac{Y^\dagger M_\Sigma^{-1} v}{\sqrt{2}}\\
 \frac{-M_\Sigma^{-1} Y v}{\sqrt{2}}(1-\frac{\epsilon}{2}) U_{\text{PMNS}} & 1-\frac{\epsilon^{\prime}}{2}
 \end{pmatrix} 
 \label{U22}
 \eea
%%%%%%%%%%%
In the above expressions $\epsilon= \frac{v^2}{2} Y^\dagger M_\Sigma^{-2} Y$, $\epsilon^\prime= \frac{v^2}{2} M_\Sigma^{-1} YY^\dagger M_\Sigma^{-1}$
and $U_{\text{PMNS}}$ is the lowest order unitary neutrino mixing matrix from \cite{Abada:2008ea,Biggio:2011ja} in accordance with \cite{Abada:2007ux}.
According to \cite{Abada:2008ea,Biggio:2011ja} $\epsilon$ and $\epsilon^\prime$ are the small quantities where we can neglect the effects of the higher powers (above $1$) of them in the calculations. For a three generation case $\mathcal{U}_L$, $\mathcal{U}_R$ and $\mathcal{V}$ are the $6\times 6$ unitary matrices. The neutrino mass can be generated by the seesaw mechanism after the diagonalization of the neutral lepton mass matrix in Eq.~\ref{n1} can be written as
\bea
m_\nu = -\frac{v^2}{2} Y^T M_\Sigma^{-1} Y
\label{mass}
\eea
%%%%%%%%%%%%%%
Due to the Eqs.~\ref{U11} and \ref{U22}, a mixing parameter between the SM lepton and the triplet is generated which affects the interaction of the triplet fermion with the SM leptons through the $W$, $Z$ and $H$ bosons. The modified charged current (CC) interaction can be written as 
\bea
 -\mathcal{L}_{\text{CC}}&=&\frac{g}{\sqrt{2}}
 \begin{pmatrix}
  \overline{e} & \overline{\Sigma} \\
 \end{pmatrix}
 \gamma^\mu W_\mu^- P_L  
 \begin{pmatrix}
 (1+\frac{\epsilon}{2}) U_{\text{PMNS}} &-\frac{Y^\dagger M_\Sigma^{-1} v}{\sqrt{2}}\\
 0&\sqrt{2}(1-\frac{\epsilon^\prime}{2})\\
 \end{pmatrix}
 \begin{pmatrix}
  \nu \\
  \Sigma^0 \\
 \end{pmatrix} \nonumber \\
 &+&\frac{g}{\sqrt{2}}
 \begin{pmatrix}
  \overline{e} & \overline{\Sigma} \\
 \end{pmatrix}
 \gamma^\mu W_\mu^- P_R  
  \begin{pmatrix}
 0&-\sqrt{2}m_{\ell} Y^\dagger M_{\Sigma}^{-2}v \\
 -\sqrt{2}M_\Sigma^{-1}Y^\ast(1-\frac{\epsilon^\ast}{2})U_{\text{PMNS}}^\ast&\sqrt{2}(1-\frac{\epsilon^{\prime^\ast}}{2})
 \end{pmatrix} 
 \begin{pmatrix}
  \nu \\
  \Sigma^0 \\
 \end{pmatrix}
  \label{CC} 
   \eea
and the modified neutral current (NC) interaction for the charged sector can be written as 
 \bea
- \mathcal{L}_{\text{NC}}^{1}&=&\frac{g}{\cos\theta_W}
 \begin{pmatrix}
  \overline{e} & \overline{\Sigma} \\
 \end{pmatrix}
 \gamma^\mu Z_\mu P_L  
 \begin{pmatrix}
  \frac{1}{2}-\cos^2\theta_W-\epsilon&\frac{Y^\dagger M_\Sigma^{-1} v}{2}\\
 \frac{M_\Sigma^{-1} Y v}{2}& \epsilon^\prime-\cos^2\theta_W
 \end{pmatrix}
 \begin{pmatrix}
  e \\
  \Sigma \\
 \end{pmatrix} \nonumber \\
 &+&\frac{g}{\cos\theta_W}
 \begin{pmatrix}
  \overline{e} & \overline{\Sigma} \\
 \end{pmatrix}
 \gamma^\mu Z_\mu P_R  
  \begin{pmatrix}
 1-\cos^2\theta_W&m_\ell Y^\dagger M_\Sigma^{-2} v \\
 M_\Sigma^{-2} Y m_{\ell} v&-\cos^2\theta_W
 \end{pmatrix} 
 \begin{pmatrix}
  e \\
  \Sigma \\
 \end{pmatrix}.
 \label{NC1} 
  \eea
The modified NC interaction for the neutral sector of the leptons can be written as 
\bea
- \mathcal{L}_{\text{NC}}^{2}&=&\frac{g}{2\cos\theta_W}
 \begin{pmatrix}
  \overline{\nu} & \overline{\Sigma^{0}} \\
 \end{pmatrix}
 \gamma^\mu Z_\mu P_L  
 \begin{pmatrix}
 1-U_{\text{PMNS}}^\dagger \epsilon U_{\text{PMNS}} & \frac{U_{\text{PMNS}}^\dagger Y^\dagger M_\Sigma^{-1} v}{\sqrt{2}}\\
 \frac{ M_\Sigma^{-1} Y_\Sigma U_{\text{PMNS}} v}{\sqrt{2}}& \epsilon^\prime
 \end{pmatrix}
 \begin{pmatrix}
  \nu \\
  \Sigma^{0} \\
 \end{pmatrix}
 \label{NC2} 
 \eea
%%%%%%%%%%%%%%%%
where $\theta_W$ is the Weinberg angle or weak mixing angle. 
The complete NC interaction can be written as $\mathcal{L}_{\text{NC}}=\mathcal{L}_{\text{NC}}^1+\mathcal{L}_{\text{NC}}^2$.
Finally we write the interaction Lagrangian of the SM leptons, charged and neutral multiplets of the  the triplet fermions with the SM Higgs $(h)$ boson.
The interaction between the charged sector and the $h$ can be written as 
%%%%%%%%%%%%%%%%%%%%%%%%%%%%%%%%%%%%%%%%%%%%%%%%%%%%
\bea
 -\mathcal{L}_{H}^1&=&\frac{g}{2M_W}
 \begin{pmatrix}
  \overline{e} & \overline{\Sigma} \\
 \end{pmatrix}
 h P_L
 \begin{pmatrix}
  -\frac{m_\ell}{v}(1-3\epsilon)& m_{\ell} Y^\dagger M_\Sigma^{-1}\\
Y(1-\epsilon)+M_\Sigma^{-2} Y m_\ell^2 & Y Y^\dagger M_\Sigma^{-1} v
 \end{pmatrix}
  \begin{pmatrix}
  e \\
  \Sigma \\
 \end{pmatrix}  \nonumber \\ 
 &+&
 \frac{g}{2M_W}
 \begin{pmatrix}
  \overline{e} & \overline{\Sigma} \\
 \end{pmatrix}
 P_R 
  \begin{pmatrix}
  -\frac{m_\ell}{v}(1-3\epsilon^\ast)& M_\Sigma^{-1}Y^\dagger m_\ell\\
(1-\epsilon^\ast) Y^\dagger+ m_\ell^2 Y_\Sigma^\dagger M_\Sigma^{-2} &  M_\Sigma^{-1}Y Y^\dagger v
 \end{pmatrix} 
   \begin{pmatrix}
  e\\
  \Sigma \\
 \end{pmatrix}
 \label{H1}
\eea
%%%%%%%%%%%%%%%%%%%%%%%%%%%%
and that same between the neutral sector and the SM Higgs can be written as
%%%%%%%%%%%%%%%%%
\bea
 -\mathcal{L}_{H}^2&=&
 \begin{pmatrix}
  \overline{\nu} & \overline{\Sigma^0} \\
 \end{pmatrix}
 h P_L
 \begin{pmatrix}
\frac{\sqrt{2} m_\nu}{v}& U_{\text{PMNS}}^T m_\nu Y^\dagger M_\Sigma^{-1} \\
(Y-\frac{Y \epsilon}{2} -\frac{\epsilon^{\prime T}Y}{2})U_{\text{PMNS}}&\frac{Y Y^\dagger M_\Sigma^{-1} v}{\sqrt{2}}
 \end{pmatrix}
  \begin{pmatrix}
  \nu \\
  \Sigma^0 \\
 \end{pmatrix}  \nonumber \\ 
 &+&
 \begin{pmatrix}
  \overline{e} & \overline{\Sigma^0} \\
 \end{pmatrix}
 P_R 
  \begin{pmatrix}
\frac{\sqrt{2} m_\nu}{v}&  M_\Sigma^{-1} Y m_\nu U_{\text{PMNS}}^\ast \\
U_{\text{PMNS}}^\ast(Y^\dagger-\frac{\epsilon^\ast Y^\dagger}{2} -\frac{Y^\dagger \epsilon^{\prime \ast}Y}{2})&\frac{M_\Sigma^{-1} Y Y^\dagger v}{\sqrt{2}}
 \end{pmatrix}
 \begin{pmatrix}
  \nu\\
  \Sigma^0 \\
 \end{pmatrix}
 \label{H2}
\eea
%%%%%%%%%%%%%
The quantities $m_\ell$ and $m_\nu$ are the SM charged lepton and tiny light neutrino mass matrices which are real and diagonal, too. 
The effect of these masses in the collider study will be negligible.
Therefore in our further analyses we do not consider the effects coming from them.
The complete Higgs interaction can be written as $\mathcal{L}_H =\mathcal{L}_H^1+\mathcal{L}_H^2$.
We want to make a comment that the general expressions calculated independently from Eq.~\ref{ml} to Eq.~\ref{H2} in this section
match exactly with the expressions given in \cite{Abada:2008ea,Biggio:2011ja}.  
The charged multiplets of the triplet fermions can interact with photons $(A_\mu)$. The corresponding Lagrangian derived from Eq.~\ref{Lm} can be written as  
%%%%%%%%%%%%
\bea
- \mathcal{L}_{\gamma \Sigma\Sigma}&=&g\sin\theta_W
 \begin{pmatrix}
  \overline{e} & \overline{\Sigma} \\
 \end{pmatrix}
 \gamma^\mu A_\mu P_L  
 \begin{pmatrix}
 1&0\\
 0&1
 \end{pmatrix}
 \begin{pmatrix}
  e \\
  \Sigma \\
 \end{pmatrix} \nonumber \\
 &+&g\sin\theta_W
 \begin{pmatrix}
  \overline{e} & \overline{\Sigma} \\
 \end{pmatrix}
 \gamma^\mu A_\mu P_R  
  \begin{pmatrix}
  1&0\\
  0&1
 \end{pmatrix} 
 \begin{pmatrix}
  e \\
  \Sigma \\
 \end{pmatrix}.
\label{Ph}
\eea
%%%%%%%%%%%%

At this point we want to mention that in the rest of the text we express the light-heavy mixing by $V_\ell = \frac{Y^Tv}{\sqrt{2} M_\Sigma}$
which can easily be obtained from the expressions given in the interactions between Eq.~\ref{CC} to Eq.~\ref{H2}.
The limit on $V_\ell$ for the electron flavor in the type-III seesaw scenario is $V_e < 0.019$ from the electroweak precision data as stated in \cite{delAguila:2008pw,delAguila:2008cj}. 
For the time being we are considering the $e^-e^+$ and $e^-p$ colliders therefore we probe $V_e$ from a variety of the final states including electron. 

\begin{figure}[]
\centering
\includegraphics[width=0.46\textwidth]{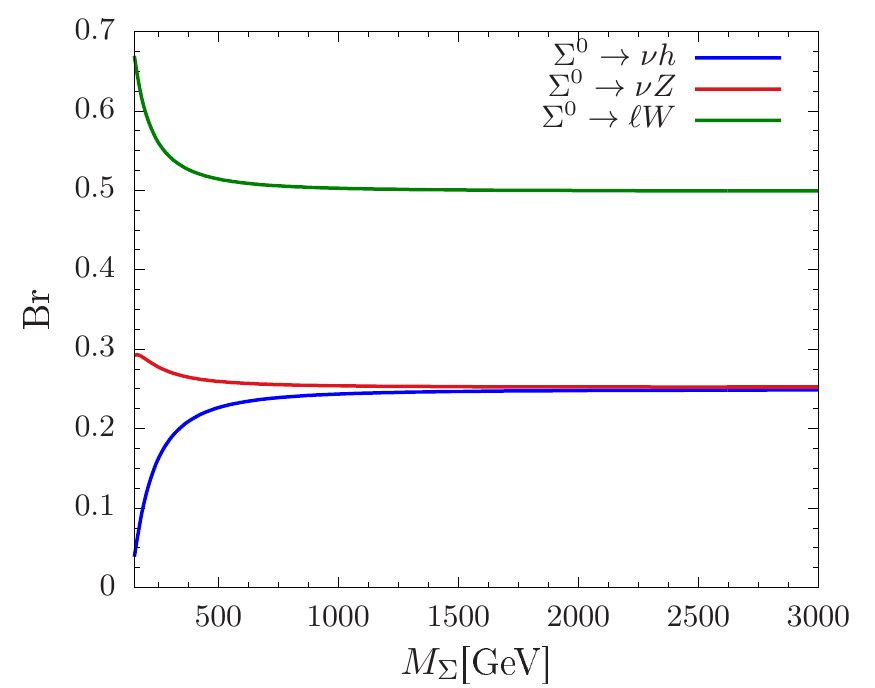}
\includegraphics[width=0.46\textwidth]{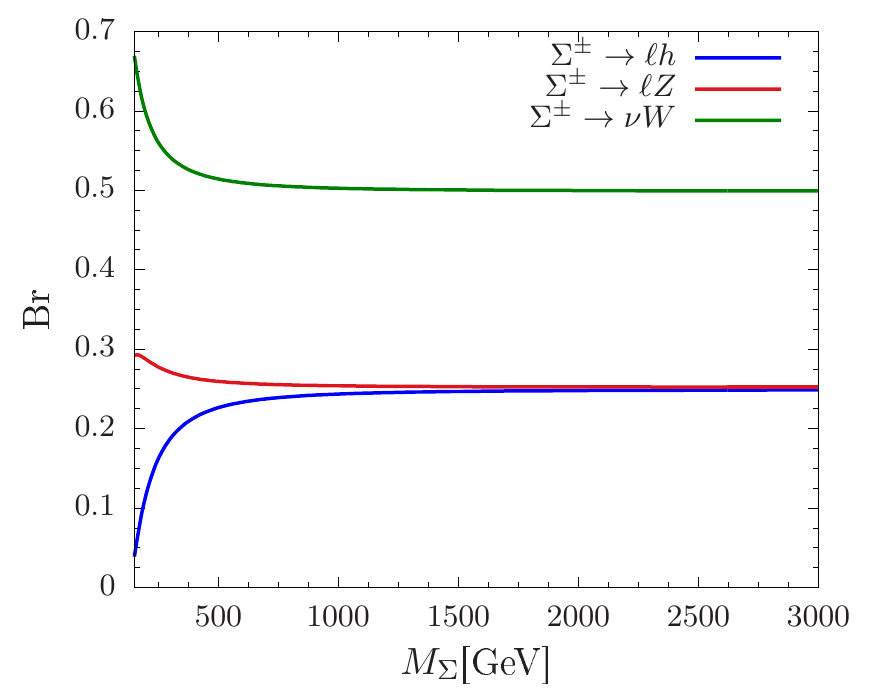}\\
\caption{Branching ratio (Br) of $\Sigma^0$~(left) and $\Sigma^{\pm}$~(right) into the SM particles as a function of $M_\Sigma$ for $V_e=0.019$, $V_\mu=0$ and $V_\tau=0$.}
\label{branching ratio1}
\end{figure}

\begin{figure}[]
\centering
\includegraphics[width=0.46\textwidth]{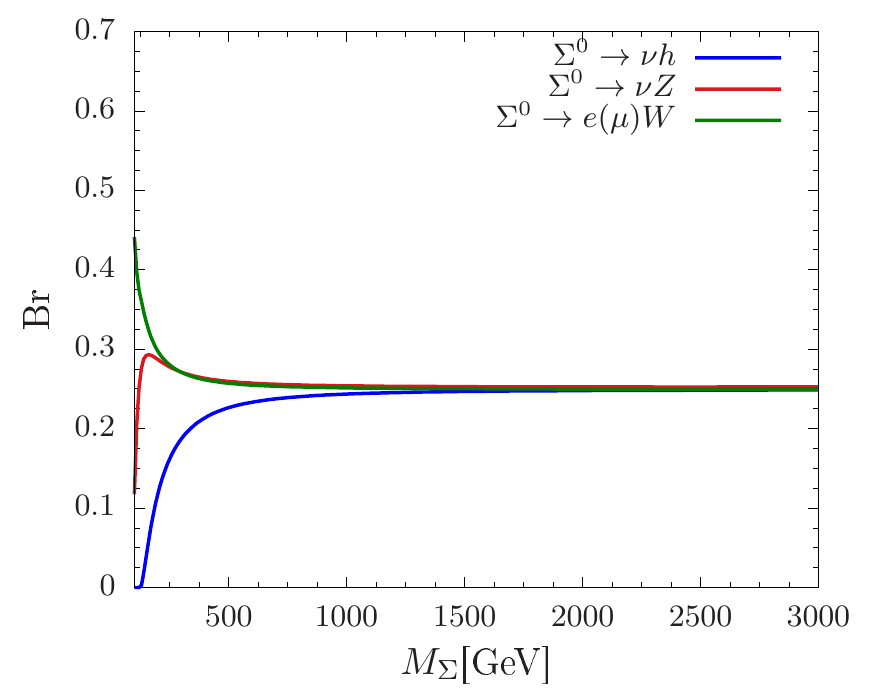}
\includegraphics[width=0.46\textwidth]{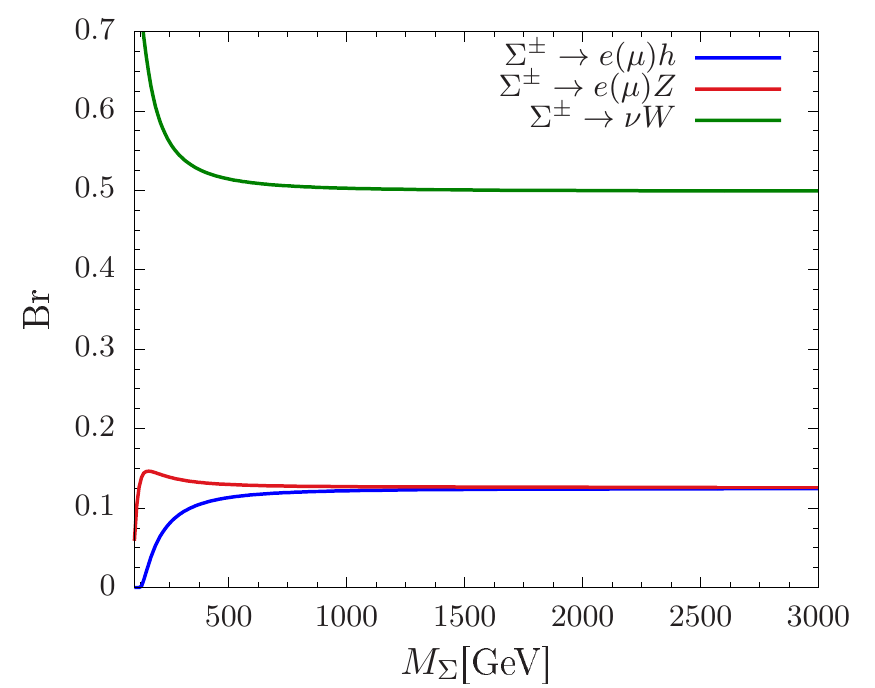}\\
\caption{Branching ratio (Br) of $\Sigma^0$~(left) and $\Sigma^{\pm}$~(right) into the SM particles as a function of $M_\Sigma$ for $V_e=V_\mu=0.0001$ and $V_\tau=0$.}
\label{branching ratio2}
\end{figure}

Using Eq.~\ref{CC} to Eq.~\ref{H2} and the expression for the mixing we calculate the partial decay widths of the neutral multiplet $(\Sigma^0)$ of the triplet fermion as
\bea
\Gamma(\Sigma^0 \to \ell^+ W)&=&\Gamma(\Sigma^0 \to \ell^- W)=\frac{g^2 |V_{\ell}|^2}{64 \pi} \Big(\frac{M_\Sigma^3}{M_W^2}\Big) \Big(1-\frac{M_W^2}{M_\Sigma^2}\Big)^2 \Big(1+2\frac{M_W^2}{M_\Sigma^2}\Big) \nonumber \\
\Gamma(\Sigma^0 \to \nu Z)&=&\frac{g^2 |V_{\ell}|^2}{64 \pi \cos^2\theta_W} \Big(\frac{M_\Sigma^3}{M_Z^2}\Big) \Big(1-\frac{M_Z^2}{M_\Sigma^2}\Big)^2 \Big(1+2\frac{M_Z^2}{M_\Sigma^2}\Big) \nonumber \\
\Gamma(\Sigma^0 \to \nu h)&=&\frac{g^2 |V_{\ell}|^2}{64 \pi} \Big(\frac{M_\Sigma^3}{M_W^2}\Big) \Big(1-\frac{M_h^2}{M_\Sigma^2}\Big)^2  
\label{decay1}
\eea
for the Majorana neutrinos and the partial decay widths of the charged multiplet $(\Sigma^\pm)$ of the triplet fermion as
\bea
\Gamma(\Sigma^\pm \to \nu W)&=&\frac{g^2 |V_{\ell}|^2}{32 \pi} \Big(\frac{M_\Sigma^3}{M_W^2}\Big) \Big(1-\frac{M_W^2}{M_\Sigma^2}\Big)^2 \Big(1+2\frac{M_W^2}{M_\Sigma^2}\Big) \nonumber \\
\Gamma(\Sigma^\pm \to \ell Z)&=&\frac{g^2 |V_{\ell}|^2}{64 \pi \cos^2\theta_W} \Big(\frac{M_\Sigma^3}{M_Z^2}\Big) \Big(1-\frac{M_Z^2}{M_\Sigma^2}\Big)^2 \Big(1+2\frac{M_Z^2}{M_\Sigma^2}\Big) \nonumber \\
\Gamma(\Sigma^\pm \to \ell h)&=&\frac{g^2|V_{\ell}|^2}{64 \pi} \Big(\frac{M_\Sigma^3}{M_W^2}\Big) \Big(1-\frac{M_h^2}{M_\Sigma^2}\Big)^2,
\label{decay2}
\eea
respectively. $M_W$, $M_Z$ and $M_h$ are the $W$, $Z$ and Higgs boson masses respectively in SM.
If the mass splitting $(\Delta M)$ between the charged $(\Sigma^\pm)$ and neutral $(\Sigma^0)$ multiplets induced by the SM gauge bosons is of the order of the pion mass \cite{Cirelli:2005uq}, $\Sigma^\pm$ can show the following additional decay modes 
\bea
\Gamma(\Sigma^\pm \to \Sigma^0 \pi^\pm)&=& \frac{2 G_F^2 V_{ud}^2 \Delta M^3 f_\pi^2}{\pi} \sqrt{1-\frac{m_\pi^2}{\Delta M^2}} \nonumber \\
\Gamma(\Sigma^\pm \to \Sigma^0 e \nu_e) &=& \frac{2 G_F^2 \Delta M^5}{15 \pi} \nonumber \\
\Gamma(\Sigma^\pm \to \Sigma^0 \mu \nu_\mu) &=&0.12 \Gamma(\Sigma^\pm \to \Sigma^0 e \nu_e)
\label{decay3}
\eea
which are independent of the free parameters. The value of the Fermi Constant, $G_F$ is $1.1663787 \times10^{-5}$ GeV$^{-2}$, the CKM matrix element $V_{ud}$ is $0.97420 \pm 0.00021$ and the $\pi$ meson decay constant, $f_\pi$, can be taken as $0.13$ GeV from \cite{Tanabashi:2018oca}. 
The Branching ratios (Br) of $\Sigma^0$ and $\Sigma^\pm$ into the SM particles are shown in left panel and right panel of Fig.~\ref{branching ratio1} as a function of $M_\Sigma$ for $V_e=0.019$, $V_\mu=0$ and $V_\tau=0$. The Branching ratios (Br) of $\Sigma^0$ and $\Sigma^\pm$ into the SM particles are shown in left panel and right panel of Fig.~\ref{branching ratio2} as a function of $M_\Sigma$ for $V_e=V_\mu=0.0001$ and $V_\tau=0$. In this paper for the further analyses we consider the case with $V_e=0.019$, $V_\mu=0$ and $V_\tau=0$ to generate the events and finally to estimate the bounds on the $|V_e|^2$.
%%%%%%%%%%%%%%%%%%%%%%%%%%
%%%%%%%%%%%%%%%%
%%%%%%%%%%%%%%%%%%%%%%%%%%%%%%%%%%%%%%%%%%%%%%%%%%%%%%%%%%%%%%
\section{Production cross-section and decay modes of the SU$(2)_L$ triplets}
\label{production}
\begin{figure}[]
\centering
\includegraphics[width=0.7\textwidth]{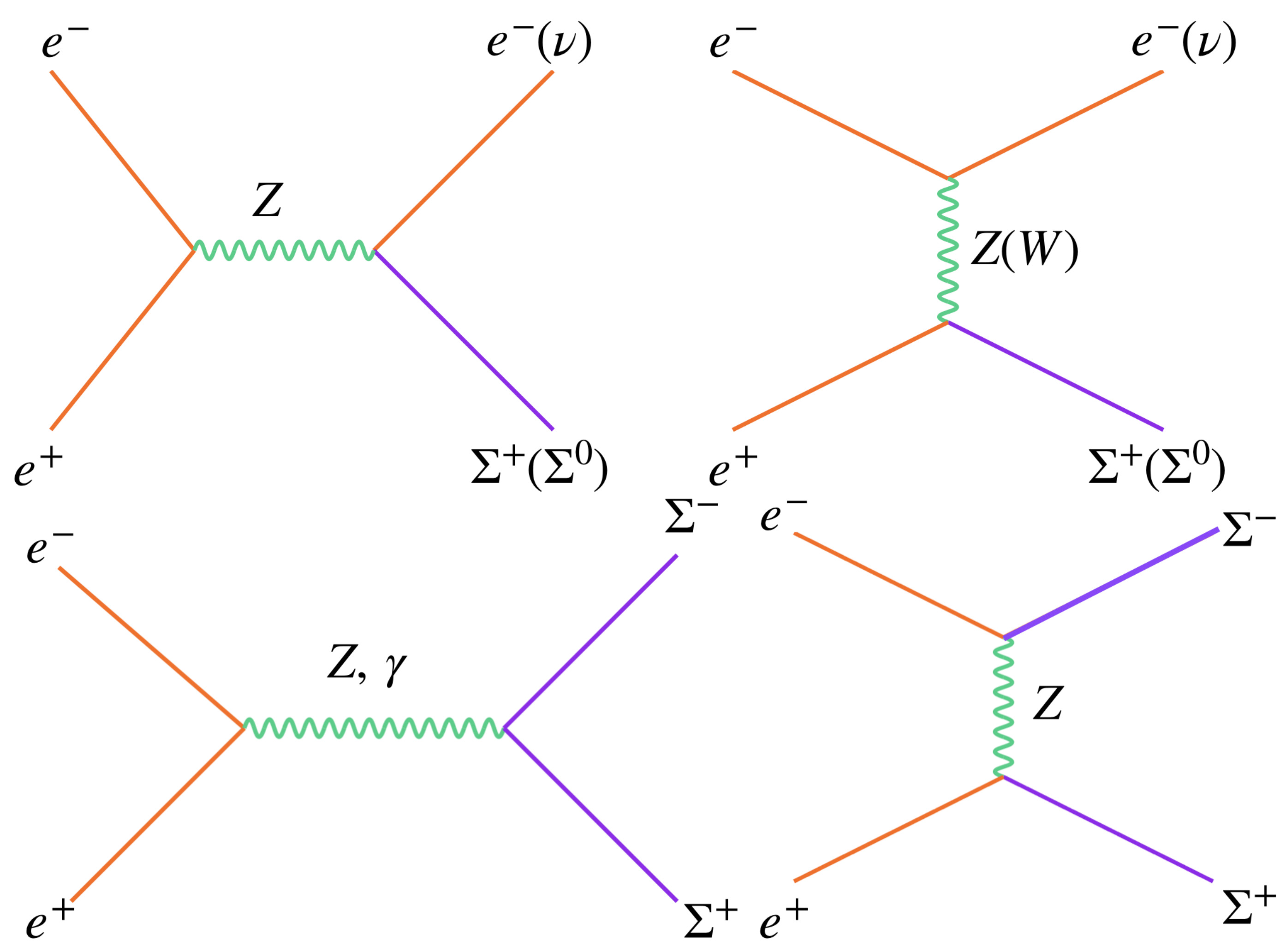}
\caption{The production modes of the $\Sigma^0$ and $\Sigma^\pm$ at the $e^-e^+$ collider.}
\label{ILC-FD-1}
\end{figure}

%%%%%%%%%%%%%%%%%%%%%%%%%%%%%%%%%%%%%%%%%%%%%%%%%%%%%%%%%%%%%
The production of the triplet fermion at the $e^-e^+$ and $e^-p$ colliders will be followed by its interactions with the SM gauge bosons as described in the Sec.~\ref{model}. 
At the $e^-e^+$ collider the triplet fermion can be produced from the $s$ and $t$ channel processes being mediated by the photon $(\gamma)$, $W$ boson and $Z$ boson respectively. The corresponding production modes are given in Fig.~\ref{ILC-FD-1}. 
%%%%%%%%%%%%%%%%%%%%%%%
%%%%%%%%%%%%%%%%%%%%%
The associate production of $\Sigma^0$ and $\Sigma^+(\Sigma^-)$ with the SM leptons (electron and neutrino) are suppressed by the light-heavy mixing square $(|V_e|^2)$. These production modes have been shown in the upper panel of the Fig.~\ref{ILC-FD-1}. At the $e^+ e^-$ collider there is another interesting production channel where $\Sigma^\pm$ is produced in pair from the $Z$ and $\gamma$ mediated processes. Such a pair-production process is direct, i. e., not suppressed by the light-heavy mixing. The corresponding production mode is given in the lower panel of Fig.~\ref{ILC-FD-1}.
%%%%%%%%%%
\begin{figure}[]
\centering
\includegraphics[width=0.49\textwidth]{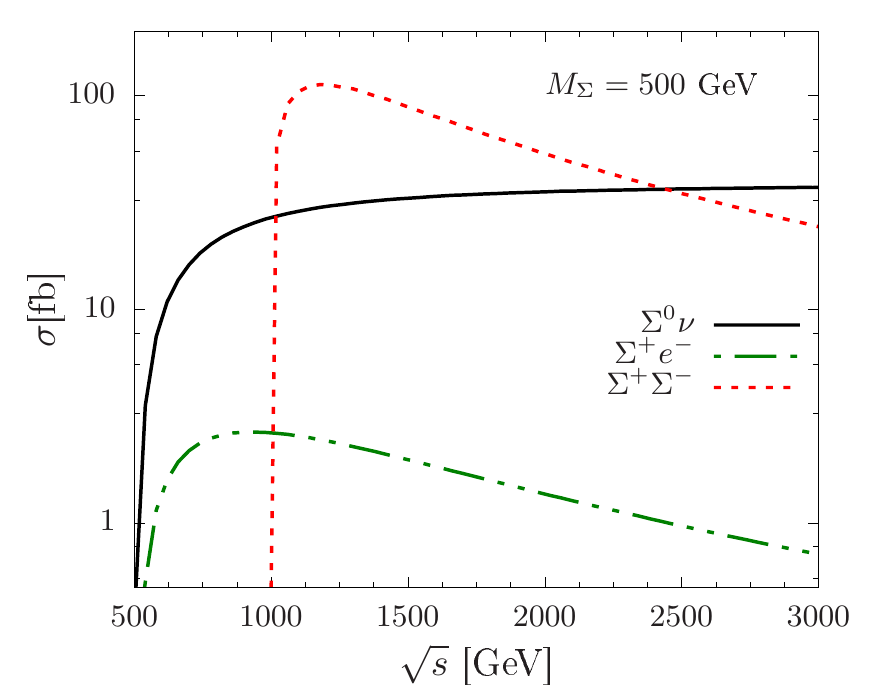}
\includegraphics[width=0.49\textwidth]{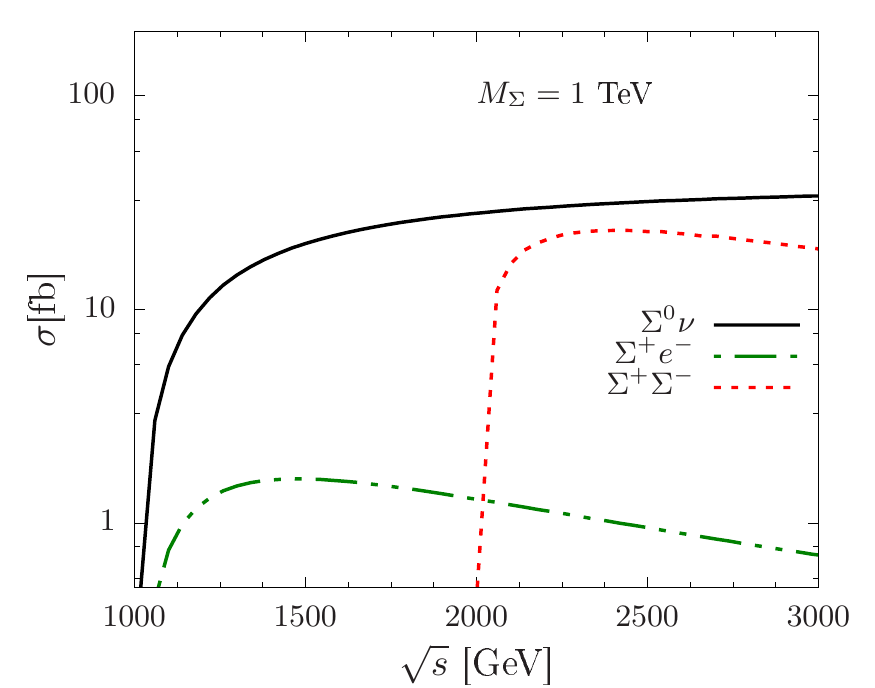}\\
\includegraphics[width=0.49\textwidth]{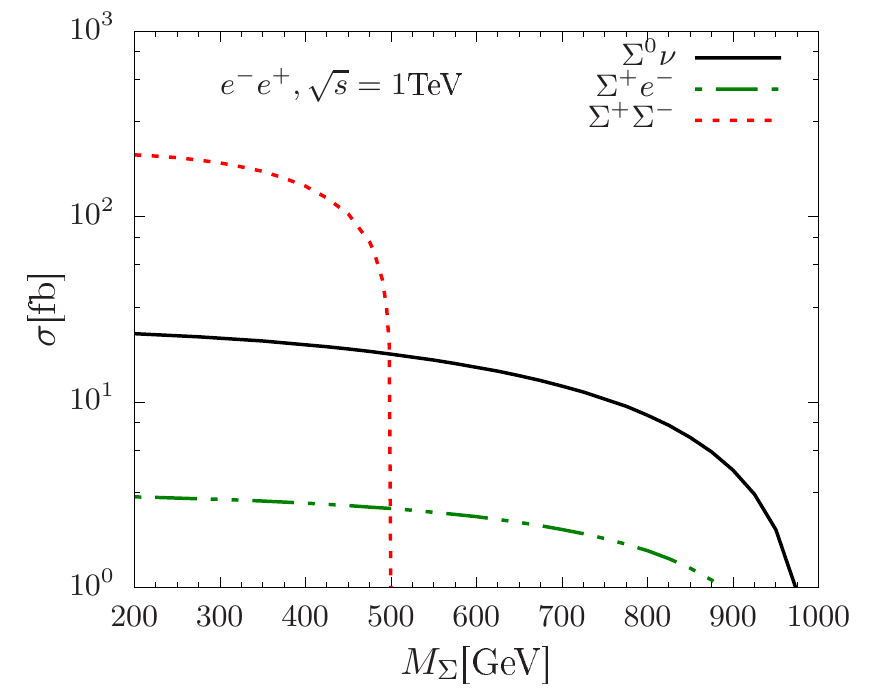}
\includegraphics[width=0.49\textwidth]{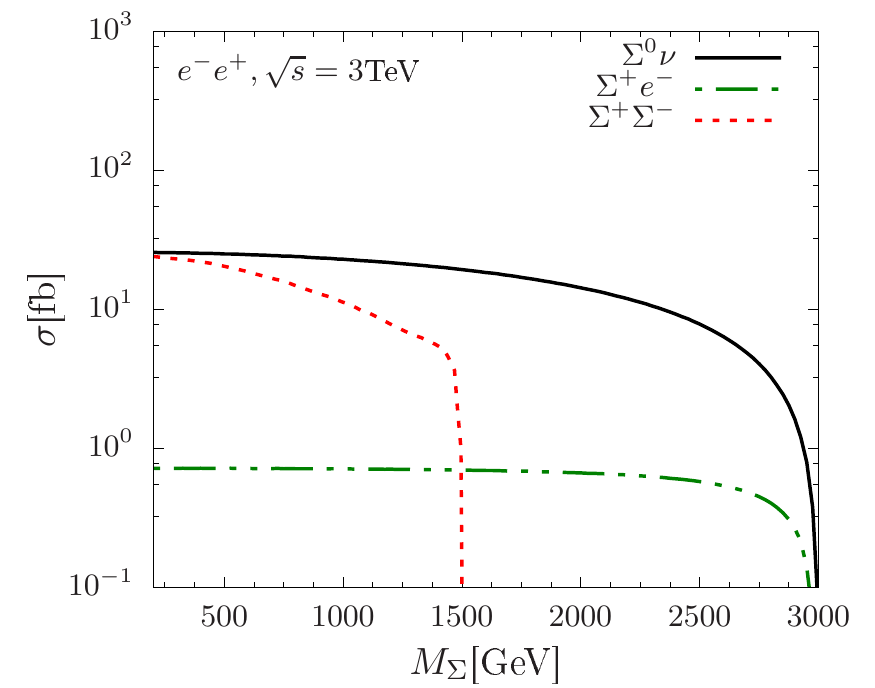}
\caption{Cross section for the triplet fermion production at the $e^-e^+$ collider. 
The production cross sections for the varying center of mass energy $(\sqrt{s})$ are shown in the upper panel 
fixing the mass of the triplet at $M_\Sigma=500$ GeV (upper, left) and at $M_\Sigma=1$ TeV (upper, right) respectively. 
The production cross sections of the triplet fermion as a function of mass are shown in the lower panel for two different values of $\sqrt{s}$ at $1$ TeV (lower, left) and $3$ TeV (lower, right), respectively. To calculate the associate production of the $\Sigma^\pm, \Sigma^0$ with the SM leptons we consider $V_e=0.019$ \cite{delAguila:2008pw,delAguila:2008cj}, however, the pair production of $\Sigma^+\Sigma^-$ is direct.}
\label{ILC-1}
\end{figure}
%%%%%%%%%%%%%%%%%%%%%%%%%%%%%%%%%

\begin{figure}[]
\centering
\includegraphics[width=0.46\textwidth]{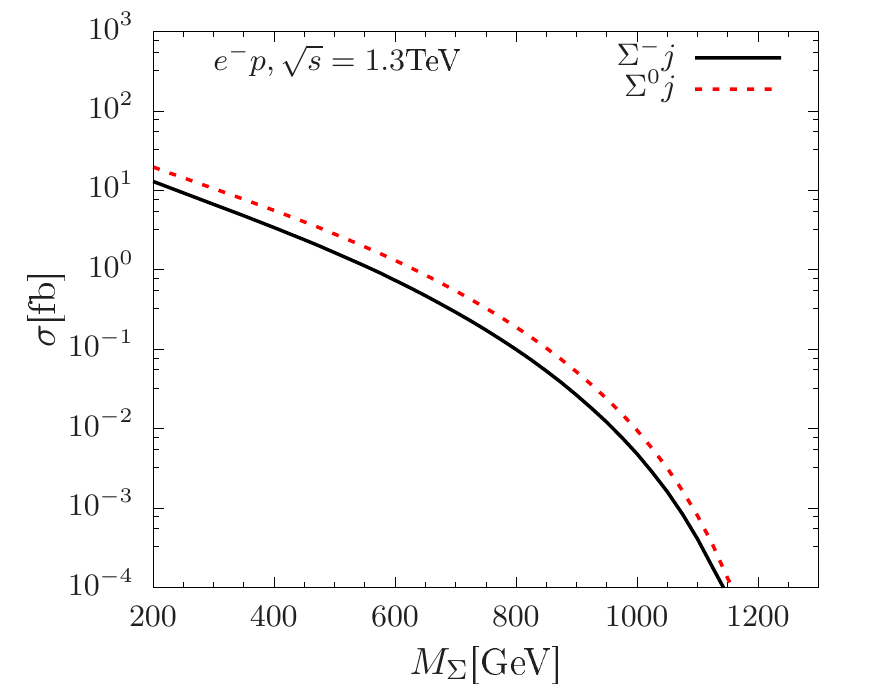}
\includegraphics[width=0.46\textwidth]{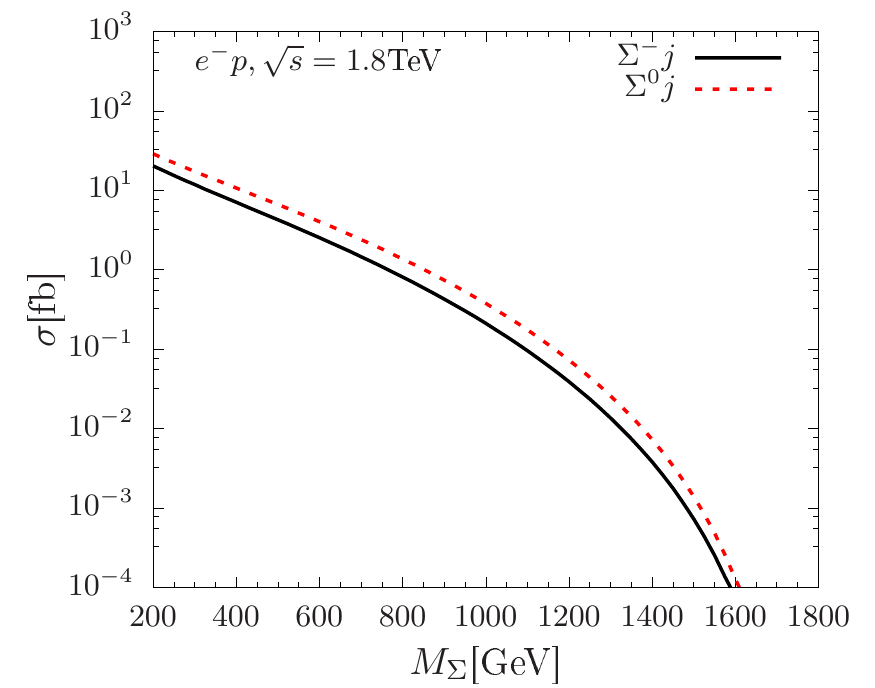}
\includegraphics[width=0.46\textwidth]{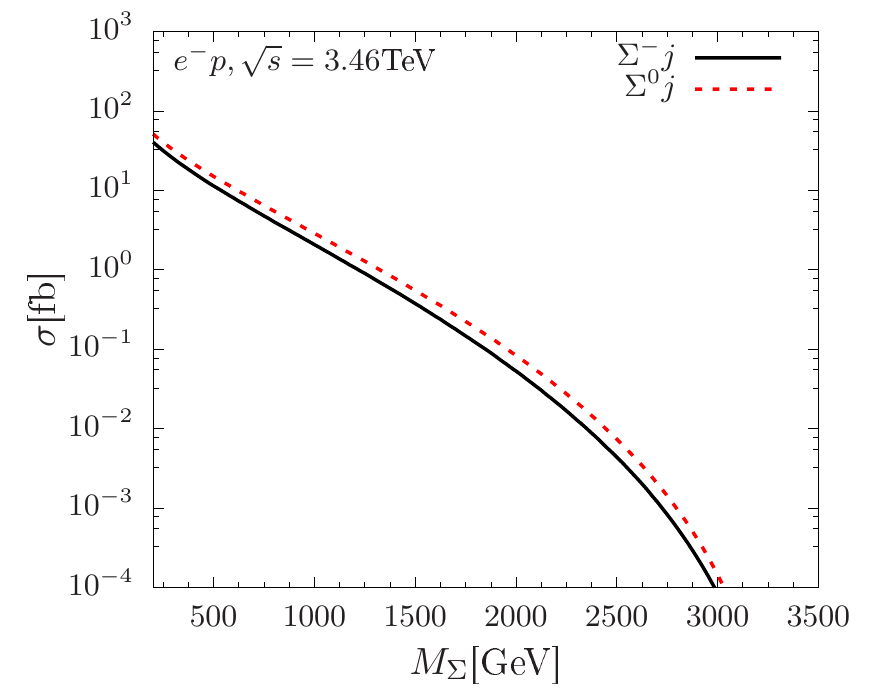}
\caption{Cross-section for the processes, $\Sigma^0 j$ and $\Sigma^- j$ in the $e^-p$ collider as a function of $M_\Sigma$ with fixed $\sqrt{s}$ at $1.3$ TeV (top, left), $1.8$ TeV (top, right) and $3.46$ TeV (bottom). We have considered $V_e=0.019$ \cite{delAguila:2008pw,delAguila:2008cj} to calculate the production cross sections.}
\label{ep-cross-section}
\end{figure}

The cross sections of different production modes of the triplet fermion at the $e^- e^+$ collider are shown in Fig.~\ref{ILC-1}. 
In the upper panel of Fig.~\ref{ILC-1} we show the production cross sections of the triplet for fixed $M_\Sigma$ but varying the center of mass energy from $500$ GeV to $3$ TeV.
In this case we fix the triplet mass at $500$ GeV (upper, left) and $1$ TeV (upper, right) respectively. In the lower panel we show the production cross section as a function of $M_\Sigma$ fixing the center of mass energy at $1$ TeV (lower, left) and $3$ TeV (lower, right) respectively.
%%%%%%%%%%%
\begin{figure}[]
\centering
\includegraphics[width=0.5\textwidth]{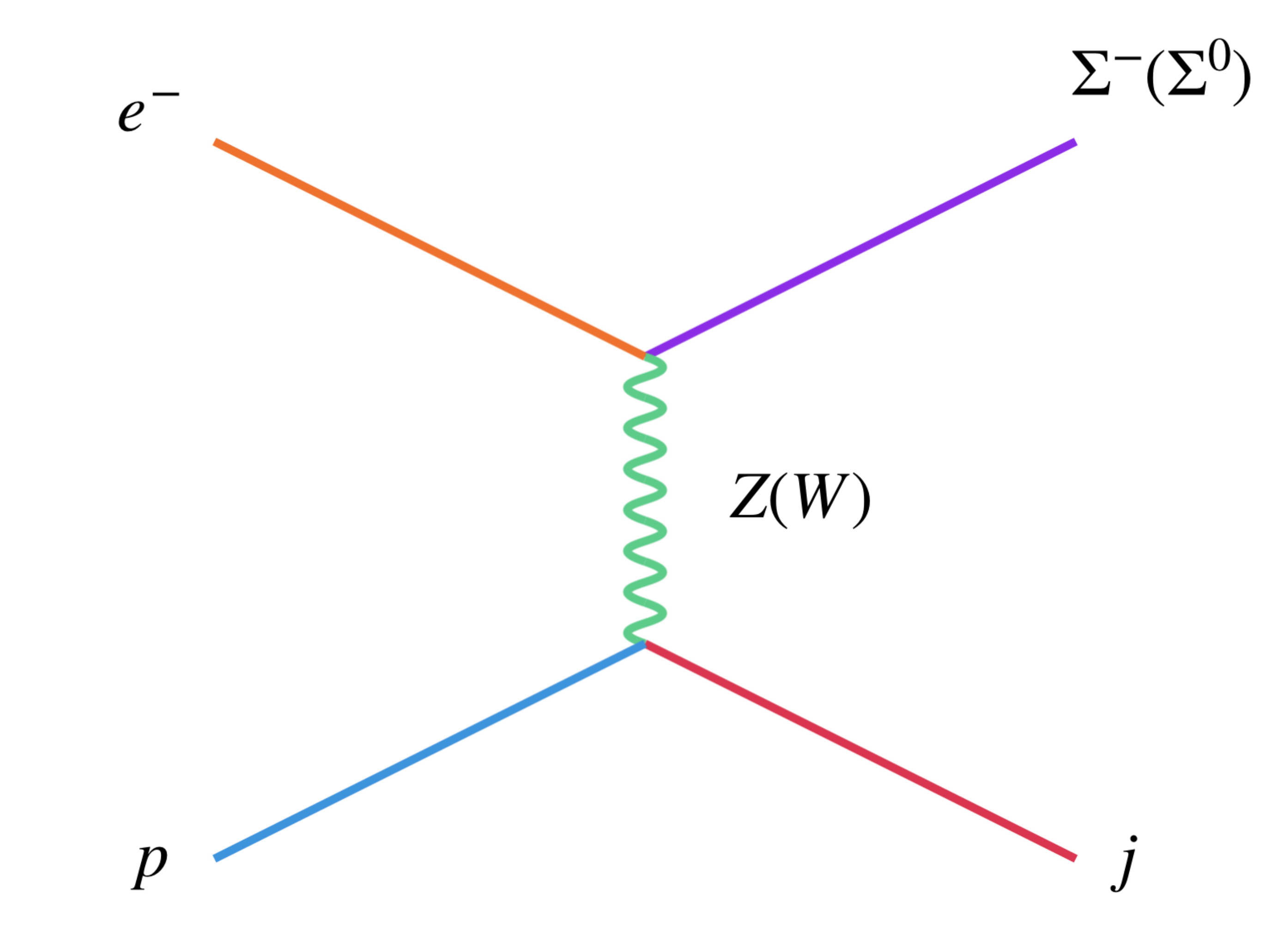}
\caption{The production modes of the $\Sigma^0$ and $\Sigma^-$ at the $e^-p$ collider which is suppressed by $|V_\ell|^2$.}
\label{LHeC-1}
\end{figure}
%%%%%%%%%%%%%%%%
The triplet fermion can be produced at the $e^-p$ collider in association with a jet through the $t$ channel process exchanging the $W$ and $Z$ bosons. The production process is shown in Fig.~\ref{LHeC-1}. We consider this process at $\sqrt{s}=1.3$ TeV, $1.8$ TeV and $3.46$ TeV, respectively. At the $e^-p$ collider the electron beam energy has been fixed at $60$ GeV but proton beam energies are $7$ TeV, $13.5$ TeV and $50$ TeV, respectively. The production processes of the $\Sigma^0 j$ and $\Sigma^- j$ are suppressed by the corresponding light-heavy mixing. The production cross sections are shown in Fig.~\ref{ep-cross-section} as a function of the $M_\Sigma$ for fixed center of mass energy $(\sqrt{s})$.
%%%%%%%%%%%%%%%%%%%%%%%%%%%%%%%%%
%%%%%%%%%%%%%%%%%%%%%%%%%%%%%%%%%%

After the production of $\Sigma^0$ and $\Sigma^\pm$ at the $e^-e^+$ and $e^-p$ colliders, the particles will decay into the respective modes according to 
Eqs.~\ref{decay1} and \ref{decay2}. In this article we consider the heavier triplet fermion so that it can sufficiently boost the decay products to form a fat jet hence we 
study the signal and the SM backgrounds. 

%%%%%%%%%%%%%%%%%%%%%%
\section{Collider Analysis}
\label{analysis}
%%%%%%%%%%%%%%%%%%%%%%%%%%%%%%%%%%%%%%%%%%%%%%%
To find the discovery prospect we first implement the model in FeynRules~\cite{Alloul:2013bka} framework, generated signal and the SM backgrounds  using the Monte Carlo event generator MadGraph5-aMCNLO~\cite{Alwall:2014hca}. For the subsequent decay, initial state radiation, final state radiation and hadronisation, we have used Pythia6~\cite{Sjostrand:2006za} for $e^-p$ and Pythia8~\cite{Sjostrand:2014zea} for $e^-e^+$ colliders respectively. We have considered the high mass regime of the triplet so that the daughter particles from triplet can be sufficiently boosted. Due to large mass gap between triplet fermion and the SM gauge bosons $(W,Z)$ and SM Higgs $(h)$, the hadronic decay modes of $W$, $Z$ and $h$ can be collimated so that we will have a single jet called fat jet $(J)$. The fat jet topology is a very powerful tool to significantly reduce the SM backgrounds. We perform the detector simulation using Delphes-3.4.1~\cite{deFavereau:2013fsa}. The detector card for the $e^-p$ collider has been obtained from~\cite{LHeCcard}. We use the ILD card for the $e^-e^+$ collider in Delphes. 
In our analysis the jets are reconstructed by Cambridge-Achen algorithm \cite{Dokshitzer:1997in,Wobisch:1998wt} implemented in Fastjet~\cite{Cacciari:2005hq,Cacciari:2011ma} package with the radius parameter as $R = 0.8$. We study the production of the triplet fermion and its subsequent decay modes at the $e^-e^+$  and $e^-p$ colliders respectively. We consider two scenarios at $e^-e^+$ collider where the center of mass energies are $\sqrt{s}=1$ TeV and $3$ TeV. For $e^-p$ collider, we consider the case of where $\sqrt{s}=3.46$ TeV. In this case the electron and proton beam energies are $60$ GeV and $50$ TeV respectively.  

At the $e^-e^+$ collider  the following set of signals after the production of the triplet fermion can be found:
\begin{enumerate}
    \item $e^-e^+\to\Sigma^0\nu~(\Sigma^\pm e^\mp)$, $\Sigma^0\to e^{\mp}W^{\mp}~(\Sigma^\pm \to \nu W^{\pm})$, $W^{\mp}\to J$, where $J$ is the fat jet coming from boosted $W$ boson. The corresponding Feynman diagram is shown in Fig.~\ref{ILC-Fat-1}.
   \item $e^-e^+\to\Sigma^0\nu~(\Sigma^\pm e^\mp)$, $\Sigma^0\to h\nu~(\Sigma^\pm \to e^\pm h)$, $h\to J_b$, where $J_b$ is the fat b-jet coming from the boosted SM Higgs decay. The corresponding Feynman diagram is shown in Fig.~\ref{ILC-Fat-2}.
   \item $e^-e^+\to\Sigma^+ e^-$, $\Sigma^+\to e^+ Z$, $Z\to J$. The corresponding Feynman diagram is shown in Fig.~\ref{ILC-Fat-3}. 
   \item $e^-e^+\to\Sigma^+\Sigma^-$, $\Sigma^\pm \to W^\pm\nu$, $W^\pm \to J$. The corresponding Feynman diagram is shown in Fig.~\ref{ILC-Fat-4}.
\end{enumerate}
At $e^-p$ collider we study the signal $e^-p\to \Sigma^0 j$ $(\Sigma^- j)$ followed by $\Sigma^0\to e^{\pm}W^{\mp}$, $W^{\mp}\to J$ $(\Sigma^0\to \nu W^{-}, W^{-}\to J)$. The corresponding Feynman diagram is shown in Fig.~\ref{LHeC-Fat-1}. As the production cross section for $\Sigma^0 j$ at $ep$ collider quickly decreases with increasing triplet mass, we decide to study the signal coming from the dominant decay mode $\Sigma^0\to eW$.

For the analyses of the signal and background events we use the following set of basic cuts:
\begin{enumerate}
\item electrons in the final state should have the following transverse momentum $(p_T^e)$ and pseudo-rapidity $(|\eta^e|)$: $p_{T}^{e}>10$ GeV, $|\eta^{e}|<2.5$~(for $ep$ collider, $|\eta^{e}|<5$).
\item jets are ordered in $p_{T}$ and they should have $p_{T}^{j}>10$ GeV and $|\eta^{j}|<2.5$~(for $e^-p$ collider, $|\eta^{j}|<5$).
\item leptons should be separated by $\Delta R_{\ell\ell}>0.2$.
\item the jets and leptons should be separated by $\Delta R_{\ell j}>0.3$.
\item fat Jet is constructed with radius parameter $R=0.8$.
\end{enumerate}
%%%%%%%%%%%%%%%%%%%%%%%%%%%%%%%%

%%%%%%%%%%%%%%%%%%%%%%%%%%%%%%%%%%%%
\subsection{Analysis for the final state $e^{\pm}+J+p_{T}^{miss}$ at $\sqrt{s}=1$ TeV and $3$ TeV $e^-e^+$ colliders}
%%%%%%%%%%%%%%%%%%%%%%%%%%%%%%%%%%%%%%%%%%%%%%%%%%%%%%%%%%%%%%
\begin{figure}[]
\centering
\includegraphics[width=0.9\textwidth]{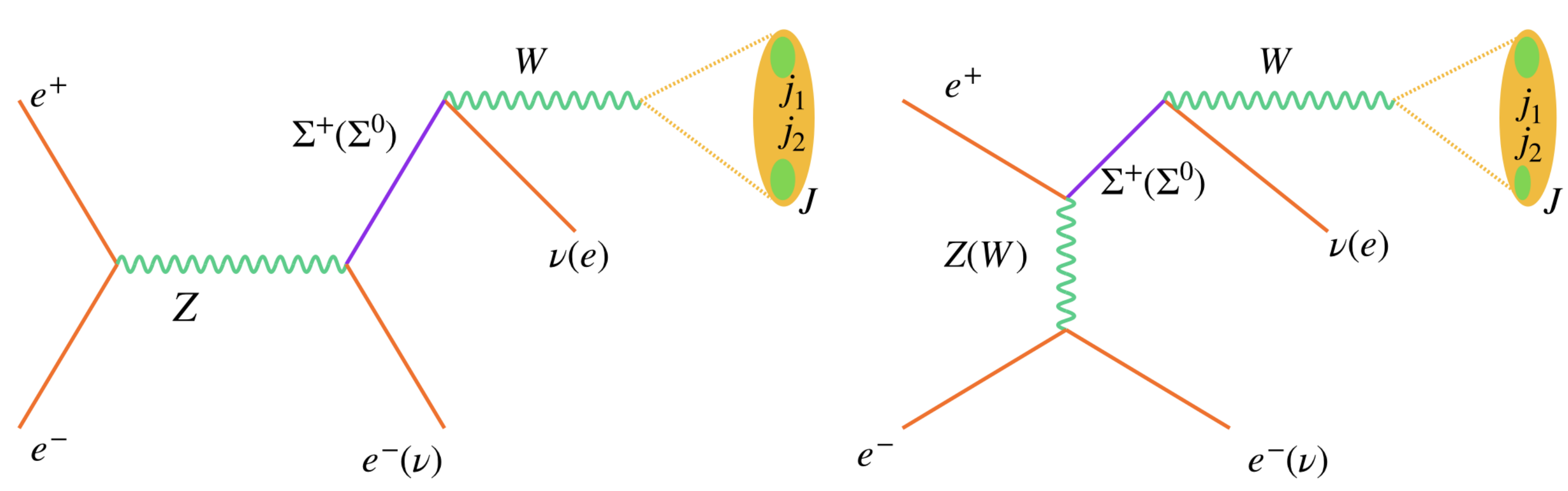}
\caption{Fat jet $(J)$ production from the $\Sigma^+$ and $\Sigma^0$ at the $e^+ e^-$ collider  from the $s$-channel (left) and $t$-channel (right) processes.}
\label{ILC-Fat-1}
\end{figure}
%%%%%%%%%%%%

The final state $e^{\pm}+J+p_{T}^{miss}$ arises from the production of $e^-e^+\to\Sigma^{0}\nu$ and the subsequent decay of $\Sigma^0$  to its dominant channel $e^{\pm}W^{\mp}$  at the $e^-e^+$ collider. The corresponding Feynman diagram is given in Fig.~\ref{ILC-Fat-1}. The $W$ boson can further decay to pair of jets. As we are considering heavy mass region of the triplet fermion, the $W$ boson will be boosted  and  its hadronic decay products, jets, will be collimated such that they can form a fat jet ($J$). 

\begin{figure}[]
\centering
\includegraphics[width=0.47\textwidth]{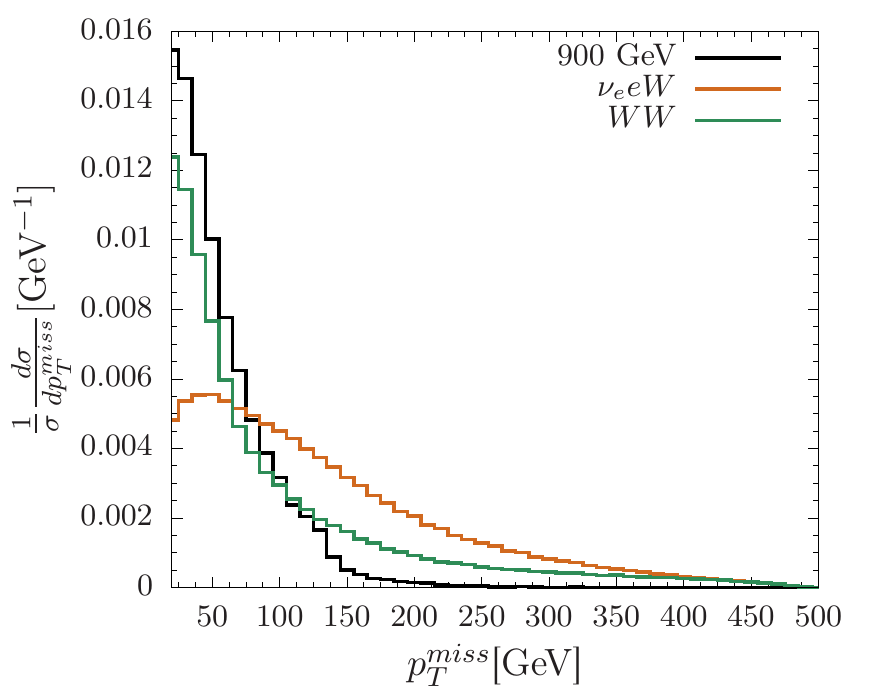}
\includegraphics[width=0.47\textwidth]{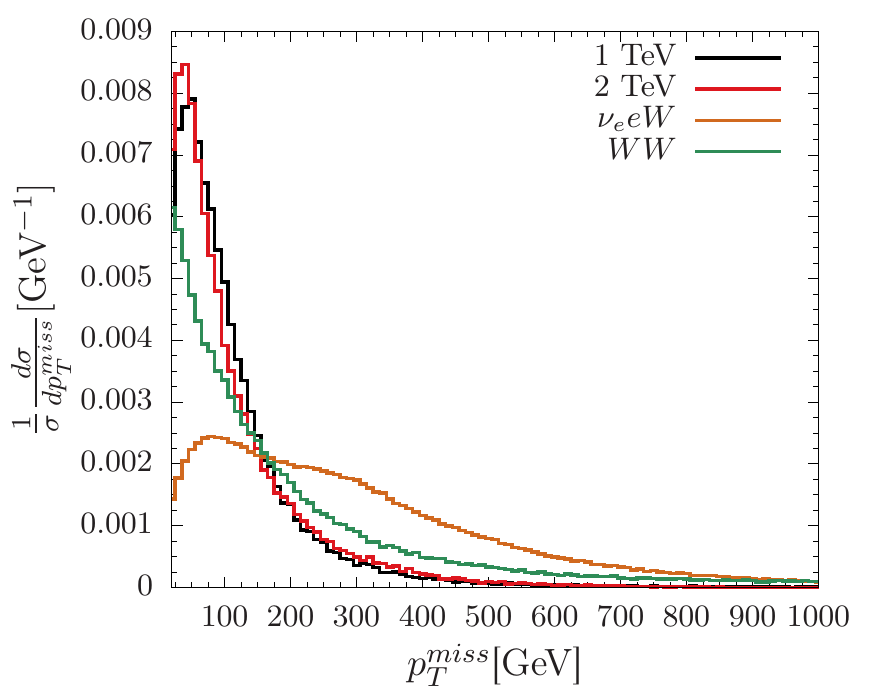}
\caption{Normalized missing momentum distributions of the signal and background events for $M_{\Sigma}$= 900 GeV at the $\sqrt{s}$= 1 TeV (left panel) and $M_\Sigma$=1, 2 TeV at the $\sqrt{s}$= 3 TeV (right panel) at $e^-e^+$ colliders.}
\label{PTMisseJMET}
\end{figure}

\begin{figure}
\includegraphics[width=0.47\textwidth]{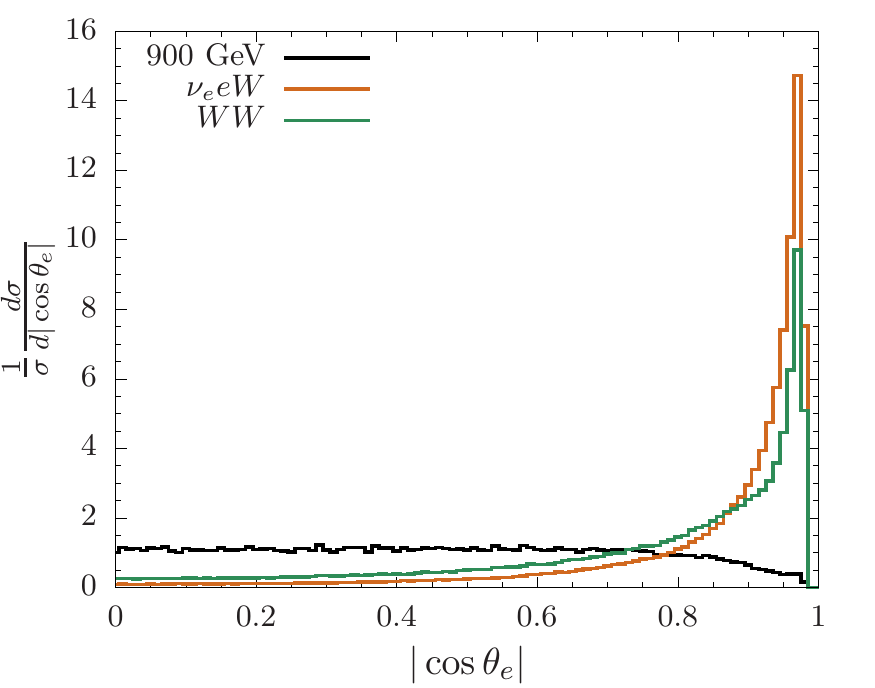}
\includegraphics[width=0.47\textwidth]{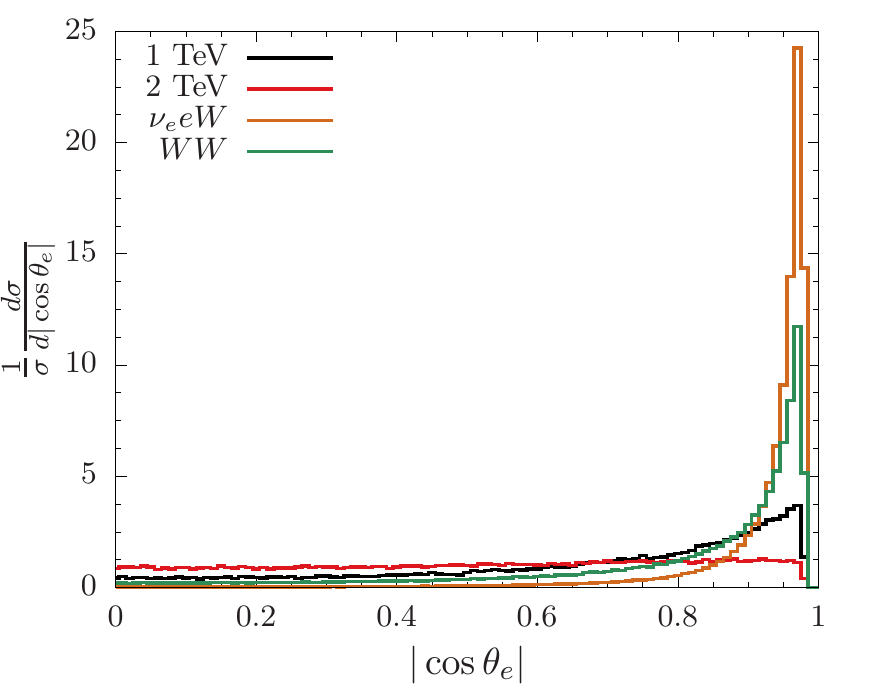}
\caption{Same as in Fig.~\ref{PTMisseJMET}, but now for $\cos\theta_e$ distribution.}
\label{costhetaleJMET}
\end{figure}

\begin{figure}
\includegraphics[width=0.47\textwidth]{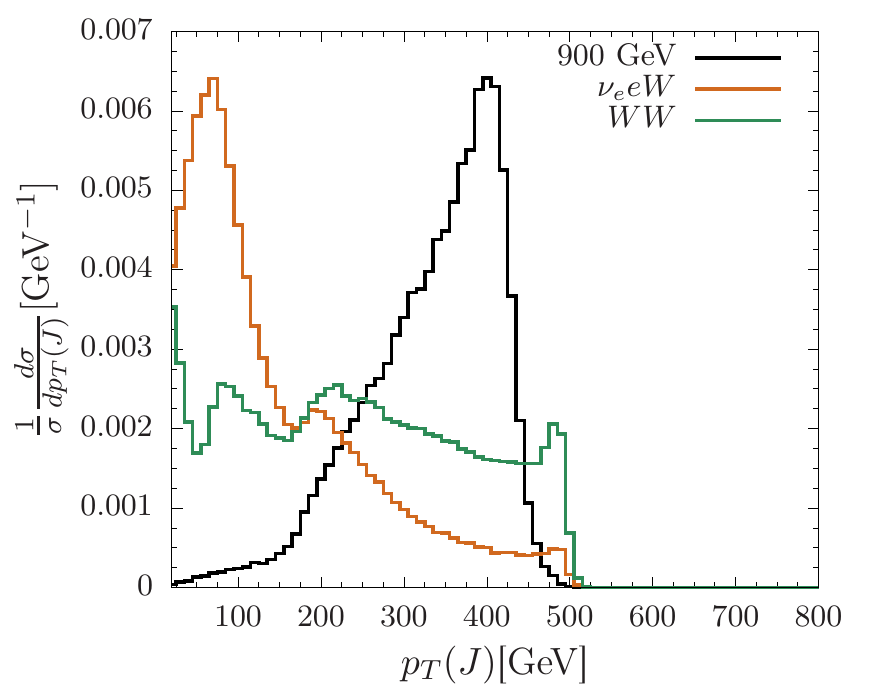}
\includegraphics[width=0.47\textwidth]{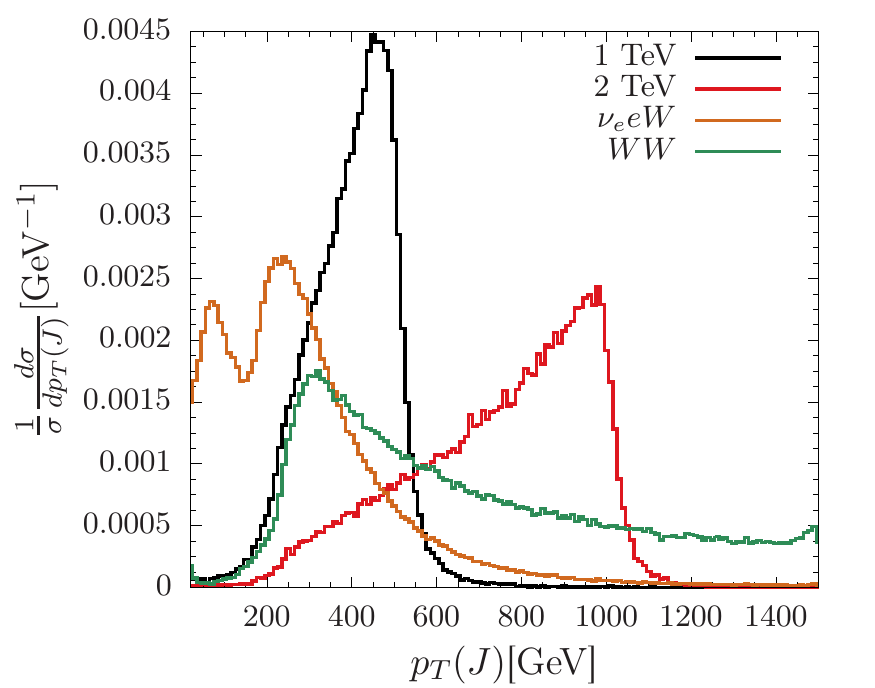}
\caption{Same as in Fig.~\ref{PTMisseJMET}, but now for fat jet $p_T$ distribution.}
\label{PTJeJMET}
\end{figure}

\begin{figure}
\includegraphics[width=0.47\textwidth]{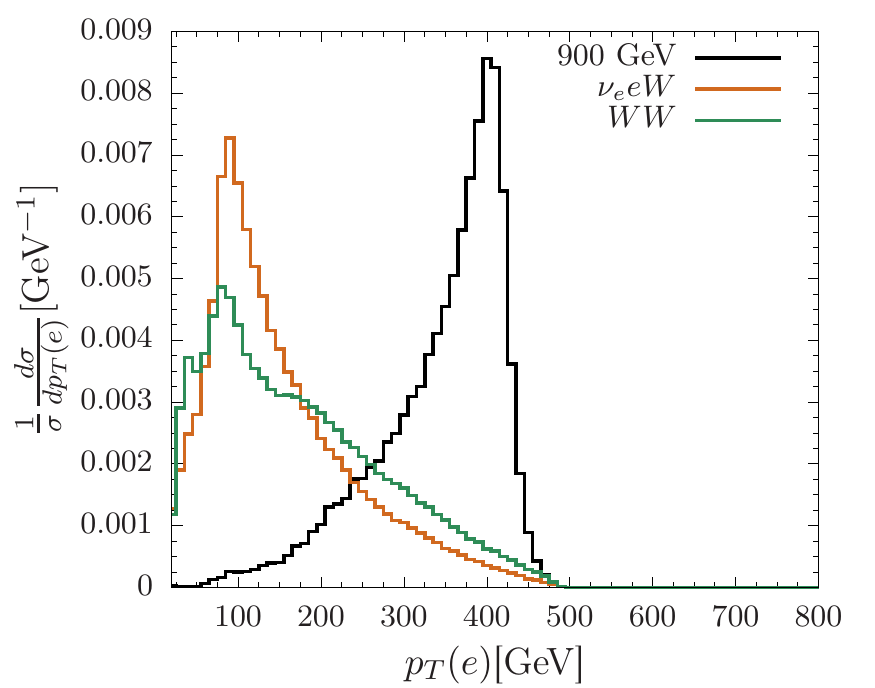}
\includegraphics[width=0.47\textwidth]{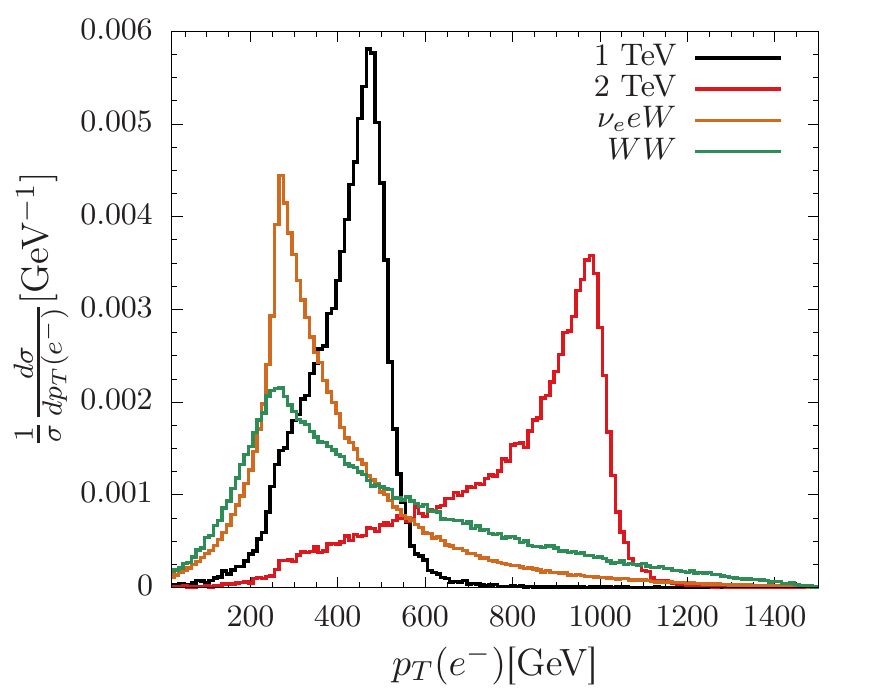}
\caption{Same as in Fig.~\ref{PTMisseJMET}, but now for leading lepton $p_T$ distribution.}
\label{PTleJMET}
\end{figure}

\begin{figure}
\includegraphics[width=0.47\textwidth]{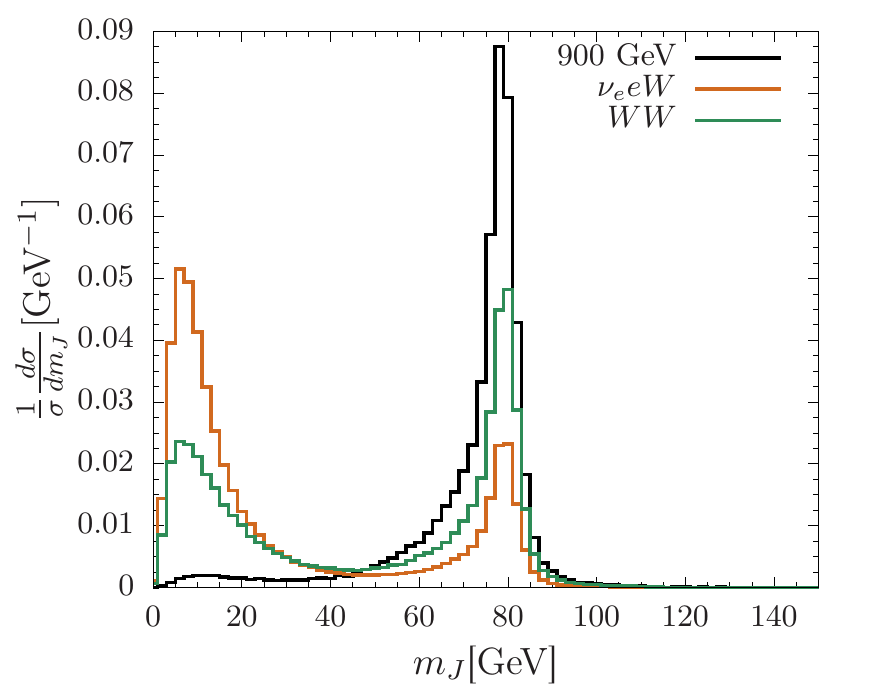}
\includegraphics[width=0.47\textwidth]{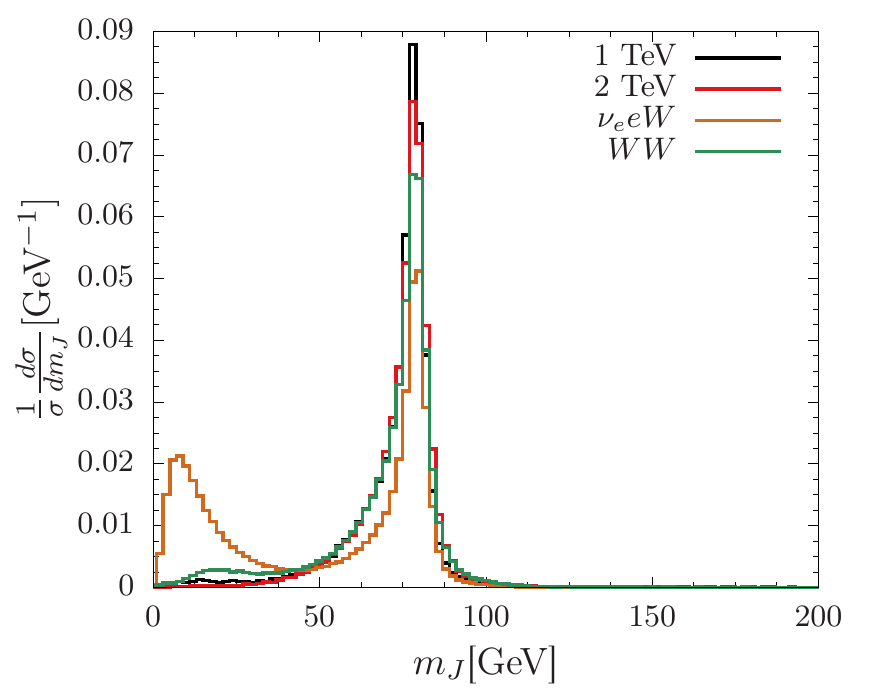}
\caption{Same as in Fig.~\ref{PTMisseJMET}, but now for invariant mass distribution of fat jet.}
\label{mJeJMET}
\end{figure}

There are a number of SM process like $\nu_e eW$, $WW$, $ZZ$ and $t\bar{t}$ which can mimic this final state which are the significant SM backgrounds. 
Among these channels $\nu_e e W$ and $WW$ give the dominant contributions. We have shown the normalized distributions of missing momentum, $|\cos\theta_e|$, fat jet $p_T$, leading lepton $p_T$ and fat jet invariant mass distributions in Figs.~\ref{PTMisseJMET}-\ref{mJeJMET} for the signal from $e^-e^+\to\Sigma^{0}\nu$ and the SM backgrounds. For these distributions we have chosen the benchmark points $M_{\Sigma}=900$ GeV at $\sqrt{s}=1$~TeV and $M_\Sigma=1,\,2$~TeV at $\sqrt{s}=3$~TeV. Note that for the case of the SM backgrounds, the invariant mass distribution of the fat jet $(m_J)$ has also low energy peaks ($m_J\leq 25$ GeV) which come from the hadronic activity of the low energy jets. Hence, a high $m_J$ cut will be useful to reduce SM backgrounds.

Due to the heavy mass of the triplet fermion, the leading lepton and the fat jet $p_T$ distributions for the signal will be in the high values than the SM backgrounds. 
Hence the high $p_T$ cut for leading lepton and fat jet will be effective to reduce SM background. At this point we would like to mention that the same final state can also be obtained from 
$e^-e^+ \to \Sigma^{\pm}e^\mp$. We have found that the $p_T$ distribution for the electron is mostly in the low momentum range for this channel. The application of the high $p_T$ 
cut for the electron applied in the $\Sigma^0 \nu$ channel will significantly cut this channel out so that its contribution becomes negligible. The high $p_T$ cut for the electron
from the $\Sigma^0 \nu$ process is required because the electron in this process is coming from the heavy triplet in contrary to the $\Sigma^\pm e^\mp$ process.
Hence in the further analyses we neglect events from the $\Sigma^{\pm}e^\mp$ process.

The polar angle variable for the electron $\cos\theta_e$ in Fig.~\ref{costhetaleJMET} is defined as $\theta_e=\tan^{-1}\big(\frac{p_T^e}{p_z^e}\big)$ where $p_z^e$ is the $z$ component of the three momentum of the electron. At the $e^-e^+$ collider the polar angle cut is very effective to reduce the SM backgrounds.

To study this process we have chosen the the triplet mass $M_{\Sigma}=800$ GeV-$950$ GeV for $\sqrt{s}=1$ TeV and $M_\Sigma=800$ GeV-$2.9$ TeV for $\sqrt{s}=3$ TeV.
In view of the distributions in Figs.~\ref{PTMisseJMET}-\ref{mJeJMET}, we have used the following set of advanced selection cuts to reduce the SM backgrounds further:
\begin{itemize}
 \item Advanced cuts for $M_{\Sigma}=800$ GeV-$900$ GeV at the $\sqrt{s}=1$ TeV $e^-e^+$ collider after the detector simulation:
\end{itemize}
%%%%%%%%%%%%%%%%%%%%%%%%%%%%%%%%%%%%%%%%%%%%
\begin{enumerate}
\item polar angle of the lepton and the fat jet $|\text{cos}~\theta_{e}|<0.9$.
\item transverse momentum for the fat jet $p_{T}^{J}>300$ GeV.
 \item transverse momentum for the leading lepton $p_{T}^{e^{\pm}}>300$ GeV.
 \item fat jet mass $m_{J}>70$ GeV.
\end{enumerate}
%%%%%%%%%%%%%%%%%%%%%%%%%%%%%%%%%%%%%%%%%%%%%%%%%%
\begin{table*}[!htbp]
	\begin{tabular}{|c|c|c|c|c|c|c|} 
		\hline
		Cuts & Signal[fb] & \multicolumn{4}{|c|}{Background[fb]} & Total [fb]\\ \hline
		&            & $\nu_{e}eW$ & $WW$ & $ZZ$ & $t\bar{t}$ &  \\ \hline
		Basic Cuts & 0.898   & 418.647 & 98.415  & 0.476     & 149.562 &  667.101 \\
		$|\text{cos}~\theta_{e}|\leq\,0.9$ & 0.863  & 165.196 & 58.901 & 0.290 & 149.551 &  373.938\\ 
		 $p^{J}_{T}>300$ GeV &  0.679 &  46.136 & 35.567 & 0.073 & 149.418 &  231.194 \\
		 $p_{T}^{e}>300\,\text{GeV}$ & 0.653     & 17.829   & 13.338      & 0.033  &  0.010 &  31.211 \\  	
		$m_{J}>70$ GeV & 0.552  & 13.905   & 10.327   & 0.027   & 0.009 &  24.269 \\
		\hline
	\end{tabular}
\caption{Cut flow for the signal and background cross-sections for the final state $e^{\pm}+J+p_T^{miss}$ for $M_{\Sigma}=900$ GeV at the $\sqrt{s}=1$ TeV $e^-e^+$ collider. 
%The signal cross section is given for the set mixing angles $V_e=0.05$, $V_\mu=0$ and $V_{\tau}=0$. 
}
\label{ILC900CM1TeV}
\end{table*}  
%%%%%%%%%%%%%%%%%%%%%%%%%%%%%%%%
%%%%%%%%%%%%%%%%%%%%%%%%%%%%%%%%%%%%%%%%%%%%
\begin{itemize}
 \item Advanced cuts for $M_{\Sigma}=800$ GeV-$2.9$ TeV at the $\sqrt{s}=3$ TeV $e^-e^+$ collider after the detector simulation:
\end{itemize}
%%%%%%%%%%%%%%%%%%%%%%%%%%%%%%%%%%%%%%%%%%%%
\begin{enumerate}
\item polar angle of the lepton and the fat jet $|\text{cos}~\theta_{e}|<0.9$.
\item transverse momentum for the fat jet $p_{T}^{J}>200$ GeV for $800$ GeV$\leq M_\Sigma \leq$$1.5$ TeV and $p_{T}^{J}>500$ GeV for  $1.6$ TeV$\leq M_\Sigma \leq 2.9$ TeV.
\item transverse momentum for the leading lepton $p_{T}^{e^{\pm}}>200$ GeV  for $800$ GeV $\leq M_\Sigma \leq 1.5 $ TeV and $p_{T}^{e^{\pm}}>500$ GeV  for $1.6$ TeV $\leq M_\Sigma \leq 2.9$ TeV.
 \item fat jet mass $m_{J}>70$ GeV.
\end{enumerate}
%%%%%%%%%%%%%%%%%%%%%%%%%%%%%%%%%
We have shown the cut flow for $M_\Sigma=900$~GeV~(at $\sqrt{s}=1$~TeV) and $M_{\Sigma}=1,\,2$ TeV (at $\sqrt{s}=3$~TeV) in Tables.~\ref{ILC900CM1TeV}-\ref{ILC2TeVCM3TeV}. Note that setting the important variable $\cos\theta_e$ as $|\cos\theta_e|\leq 0.9$ puts a very strong cut for the SM backgrounds. 
\begin{table*}[!htbp]
	\begin{tabular}{|c|c|c|c|c|c|c|} 
		\hline
		Cuts & Signal[fb] & \multicolumn{4}{|c|}{Background[fb]} & Total [fb]\\ \hline
		&            & $\nu_{e}eW$ & $WW$ & $ZZ$ & $t\bar{t}$ &  \\ \hline
		Basic Cuts & 4.611   & 267.680 & 12.686  & 0.138     & 0.182 & 280.687 \\
		$|\text{cos}~\theta_{e}|\leq\,0.9$ & 3.418  & 49.100 & 6.198 & 0.059 & 0.150 & 55.508 \\ 
		 $p^{J}_{T}>200$ GeV &  3.318 &  38.553 & 6.165 & 0.057 & 0.146 &  44.921\\
		 $p_{T}^{e}>200\,\text{GeV}$ & 3.265     & 36.982   & 5.773      & 0.047  & 0.083  &  42.886\\  	
		$m_{J}>70$ GeV & 2.499  & 27.159   & 4.484   & 0.040   & 0.077 & 31.760 \\
		\hline
	\end{tabular}
\caption{Cut flow for the signal and background cross-sections for the final state $e^{\pm}+J+p_T^{miss}$ for $M_{\Sigma}=1$ TeV at the $\sqrt{s}=3$ TeV $e^-e^+$ collider. 
%The signal cross section is given for the set of mixing angles $V_e=0.05$, $V_\mu=0$ and $V_{\tau}=0$. 
}
\label{ILC1TeVCM3TeV}
\end{table*}  
%%%%%%%%%%%%%%%%%%%%%%%%%%%%%%%%%%%%%%%%%%%%%
\begin{table*}[!htbp]
	\begin{tabular}{|c|c|c|c|c|c|c|} 
		\hline
		Cuts & Signal[fb] & \multicolumn{4}{|c|}{Background[fb]} & Total [fb]\\ \hline
		&            & $\nu_{e}eW$ & $WW$ & $ZZ$ & $t\bar{t}$ &  \\ \hline
		Basic Cuts & 2.751   & 267.680 & 12.686  & 0.138     & 0.182 & 280.687 \\
		$|\text{cos}~\theta_{e}|\leq\,0.9$ & 2.474  & 49.100 & 6.198 & 0.059 & 0.150 & 55.508 \\ 
		 $p^{J}_{T}>500$ GeV &  2.250 &  19.555 & 5.948 & 0.053 & 0.131 &  25.687\\
		 $p_{T}^{e}>500\,\text{GeV}$ & 2.223     & 15.591   & 4.346      & 0.034  & 0.043  & 20.015 \\  	
		$m_{J}>70$ GeV & 1.802  & 12.897   & 3.473   & 0.029   & 0.042 & 16.441 \\
		\hline
	\end{tabular}
\caption{Cut flow for the signal and background cross-sections for the final state $e^{\pm}+J+p_T^{miss}$ for $M_{\Sigma}=2$ TeV at the $\sqrt{s}=3$ TeV $e^-e^+$ collider. 
%The signal cross section is given for the set of mixing angles $V_e=0.05$, $V_\mu=0$ and $V_{\tau}=0$. 
}
\label{ILC2TeVCM3TeV}
\end{table*}  
To study the heavier triplet fermion at the $\sqrt{s}=3$ TeV $e^-e^+$ collider we have chosen stronger cuts for the transverse momenta of the lepton and fat jet to reduce the SM backgrounds.

%%%%%%%%%%%%%%%%%%%%%%%%%%%%%%%%%%%%%%%%%%%%%%
\subsection{Analysis for the final state $J_{b}+p_{T}^{miss}$ at $\sqrt{s}=3$ TeV $e^-e^+$ collider}
%%%%%%%%%%%%%%%%%%%%%%%%%%%%%%%%%%%%
\begin{figure}[]
\centering
\includegraphics[width=0.95\textwidth]{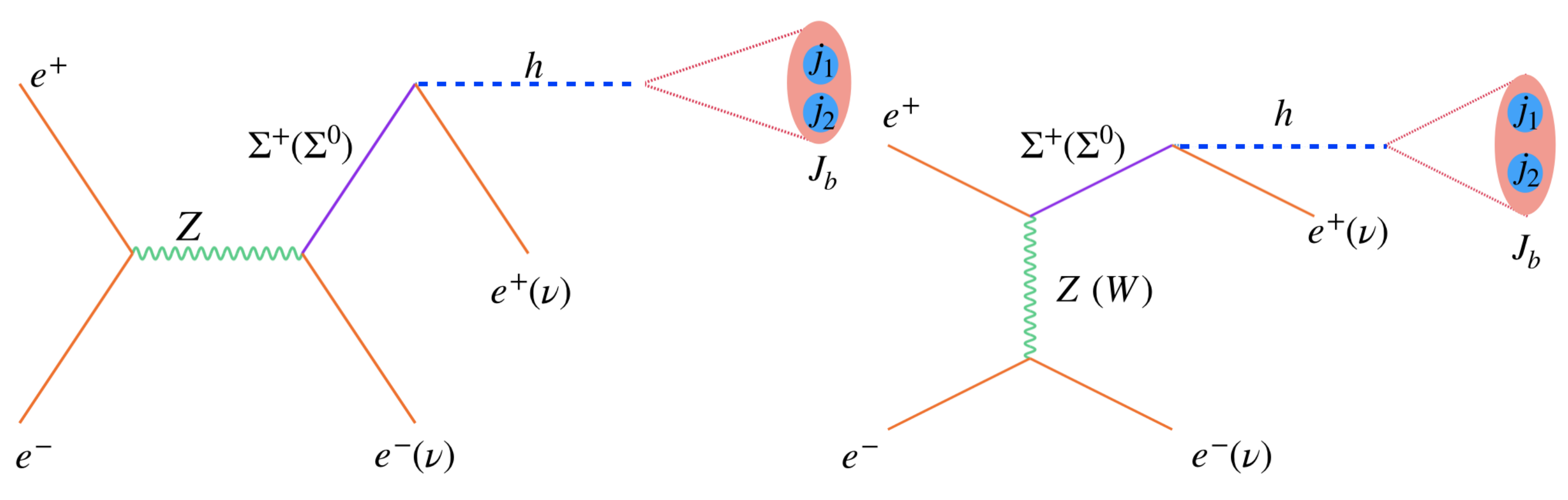}
\caption{Fat b-jet $(J_b)$ production from the $\Sigma^+$ and $\Sigma^0$ at the $e^- e^+$ collider  from the $s$-channel (left) and $t$-channel (right) processes.}
\label{ILC-Fat-2}
\end{figure}
This final state $J_{b}+p_{T}^{miss}$ arises from the production of $e^-e^+\to\Sigma^{0}\nu$ (conjugate process implied) and the subsequent decay $\Sigma^0$ to $h\nu$ at the $e^-e^+$ collider. The corresponding Feynman diagram is given in Fig.~\ref{ILC-Fat-2}. The SM Higgs $(h)$ branching ratio is $(\sim 60\%)$ to $b\bar{b}$ at $m_h=125$ GeV which is the reason for our consideration of this channel.  As $h$ is boosted in our case, we will have a collimated fat-b jet. For this final state the dominant SM backgrounds come from the process $h\nu_\ell\bar{\nu_\ell}$ and $Z\nu_\ell\bar{\nu_\ell}$. Backgrounds can also come from the processes like $Zh$ and $ZZ$, with subsequent decays of of the $Z$ boson into the light neutrinos and $h\to b \overline{b}$. We have combined all the SM backgrounds at the time of the event generation. In this work, we consider a flat $70\%$ tagging efficiency for each of the daughter $b$ jets coming from the Higgs decay. 
\begin{figure}[]
\centering
\includegraphics[width=0.47\textwidth]{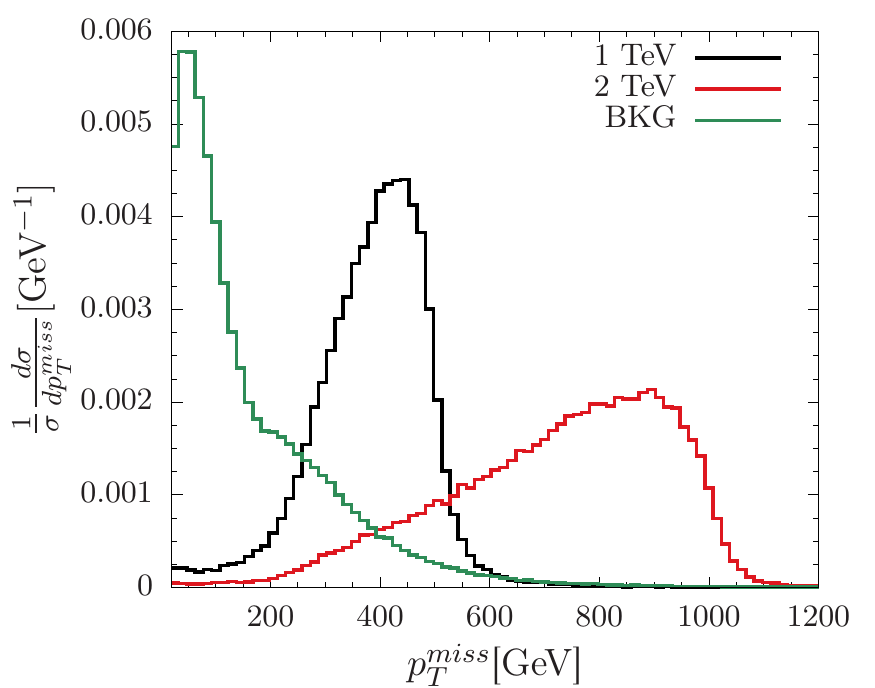}
\includegraphics[width=0.47\textwidth]{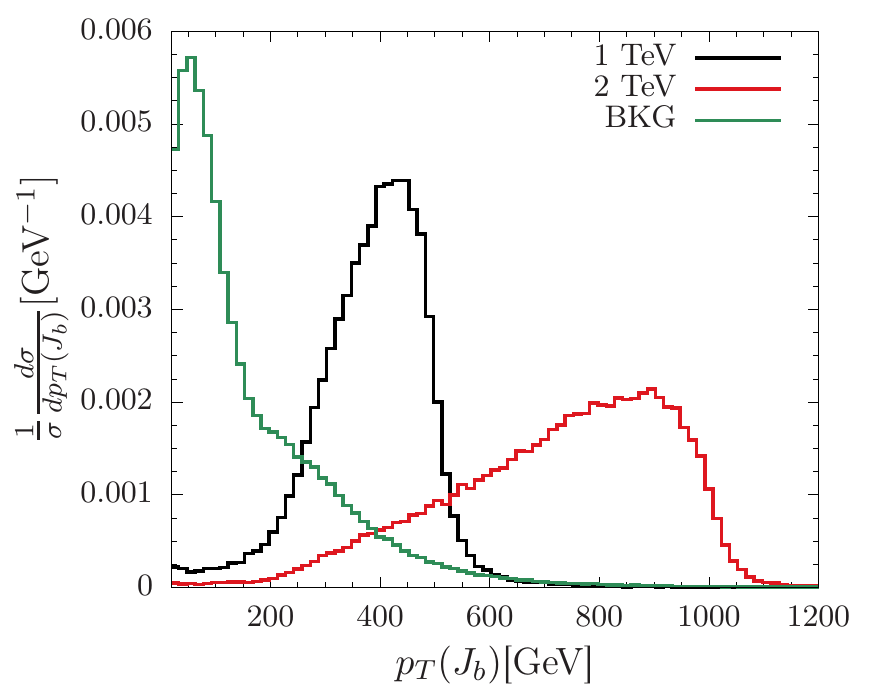}
\includegraphics[width=0.47\textwidth]{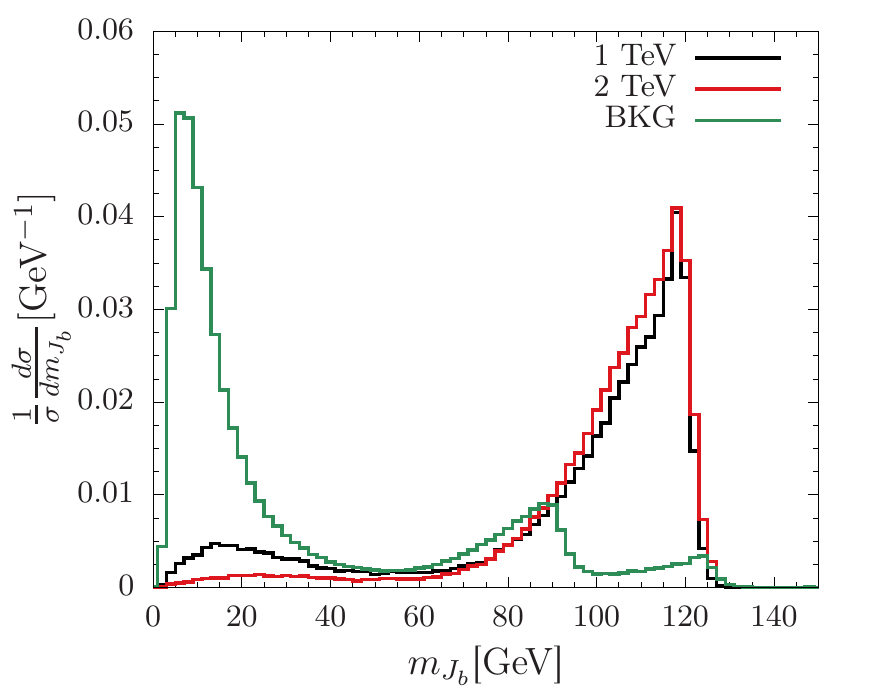}
\caption{Normalized distributions of missing momentum, fat-b $p_T$ and fat-b invariant mass of the signal and background events for $M_\Sigma$=1, 2 TeV at the $\sqrt{s}$= 3 TeV at $e^-e^+$ colliders.}
\label{ILC3TeVHistFatbMET}
\end{figure}

We have shown the normalized distributions for missing momentum, $p_T$ of the fat-b and invariant mass $m_{J_b}$ in Fig.~\ref{ILC3TeVHistFatbMET}. These distributions are given for $M_\Sigma=1,\,2$~TeV at $\sqrt{s}=3$~TeV $e^-e^+$ collider along with SM backgrounds. The invariant mass distribution of the fat-b coming from the $h$ decay peaks around the Higgs mass for the signal. Hence a cut like $m_{J_b}>100$~GeV sets a strong constraint on SM backgrounds. Missing momentum and the $p_T$ distribution of the fat-b for the signal will likely be in the high values compared to the SM backgrounds due to high mass of triplet fermion. We have considered the mass range $M_\Sigma=800$ GeV-$2.9$ TeV for $\sqrt{s}=3$~TeV. In view of these distributions of Fig.~\ref{ILC3TeVHistFatbMET}, we impose the following set of advanced selection cuts to reduce the SM backgrounds further:
%%%%%%%%%%%%%%%%%%%%%%%%%%%%%%%%%%%%%%
\begin{enumerate}
\item missing energy, $p_{T}^{miss}>300$ GeV for $800$ GeV $\leq M_{\Sigma}\leq 1.5$ TeV and $p_{T}^{miss}>500$ GeV for $1.6$ TeV $\leq M_{\Sigma}\leq$ 2.9 TeV. 
\item transverse momentum for  $J_b$, $p_{T}^{J_b}>300$ GeV for $800$ GeV $\leq M_{\Sigma}\leq 1.5$ TeV and $p_{T}^{J_b}>500$ GeV for $1.6$ TeV $\leq M_{\Sigma}\leq 2.9$ TeV.
\item fat-b mass, $m_{J_b}>100$ GeV.
\end{enumerate}
%%%%%%%%%%%%%%%%%%%%%%%%%%%%
\begin{table*}[!htbp]
 \begin{tabular}{|c|c|c|}
  \hline
  Cuts & Signal[fb]  & Background[fb] \\ \hline
  \hline
  Basic Cuts & 1.203 & 126.225 \\   
   $p_{T}^{miss}>300$ GeV & 1.007  & 21.664 \\
  $p_{T}^{J_{b}}>300$ GeV & 1.004  & 21.449 \\
  $m_{J_{b}}>100$ GeV & 0.713  & 6.470 \\ 
  \hline
 \end{tabular}
\caption{Cut flow for the signal and background cross sections for the final state $J_b+ p_T^{miss}$ for $M_{\Sigma}=1$ TeV at the $\sqrt{s}=3$ TeV linear collider. 
%The signal events are for the set of mixing angles $V_e=0.05$, $V_\mu=0$ and $V_\tau=0$. 
}
\label{ILCHiggs1TeV}
\end{table*}
%%%%%%%%%%%%%%%
\begin{table*}[!htbp]
 \begin{tabular}{|c|c|c|}
  \hline
  Cuts & Signal[fb]  & Background[fb] \\ \hline
  \hline
  Basic Cuts & 0.887 & 126.225 \\   
   $p_{T}^{miss}>500$ GeV & 0.757  & 5.025 \\
  $p_{T}^{J_{b}}>500$ GeV & 0.756  & 5.004 \\
  $m_{J_{b}}>100$ GeV & 0.571  & 1.501 \\ 
  \hline
 \end{tabular}
\caption{Cut flow for the signal and background cross sections for the final state $J_b+ p_T^{miss}$ for $M_{\Sigma}=2$ TeV at the $\sqrt{s}=3$ TeV $e^-e^+$ collider. 
%The signal events are for mixing $V_e=0.05$. 
}
\label{ILCHiggs2TeV}
\end{table*}
We have shown the cut flow for two benchmark points $M_{\Sigma}=1$ TeV and 2 TeV at $\sqrt{s}=3$~TeV in Table.~\ref{ILCHiggs1TeV} and \ref{ILCHiggs2TeV}, respectively. 
%%%%%%%%%%%%%%%%%%%%%%%%%%%%%%%%%%%%%%%%%%%%%%%%%%%%%%%%%%%
\subsection{Analysis for the final state  $e^- e^+ + J$ at $\sqrt{s}=3$ TeV $e^-e^+$ collider}
%%%%%%%%%%%%%%%%%%%%%%%%%%%%%%%%%%%%%%%%%%%%%%%%%%%%%%%%%%%%%%%%%%%%%%%%%
\begin{figure}[]
\centering
\includegraphics[width=0.9\textwidth]{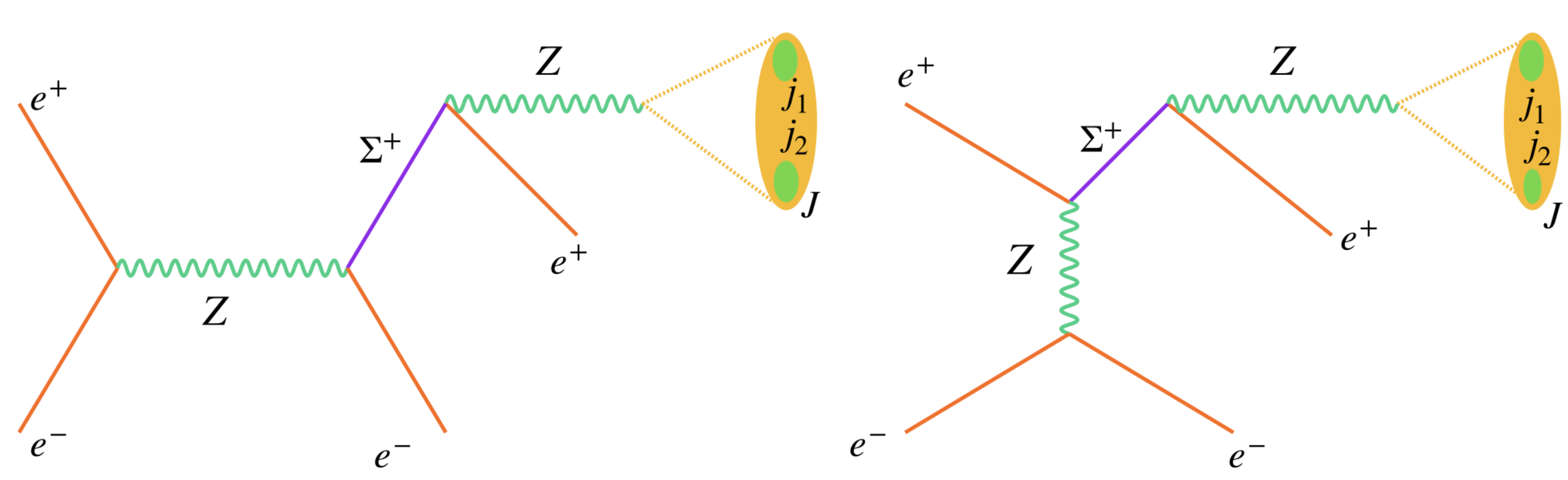}
\caption{Fat jet $(J)$ production from the $\Sigma^+$ at the $e^-e^+$ collider from the $s$-channel (left) and $t$-channel (right) processes.}
\label{ILC-Fat-3}
\end{figure}
%%%%%%%%%%%%%%%%%%%%%%%%%%%%%%%

\begin{figure}[]
\centering
\includegraphics[width=0.46\textwidth]{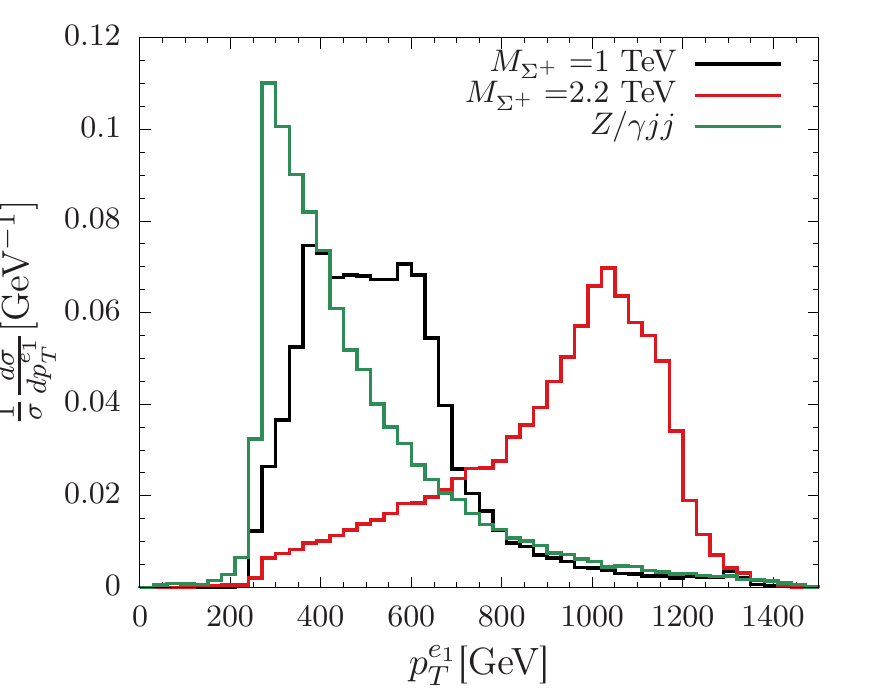}
\includegraphics[width=0.46\textwidth]{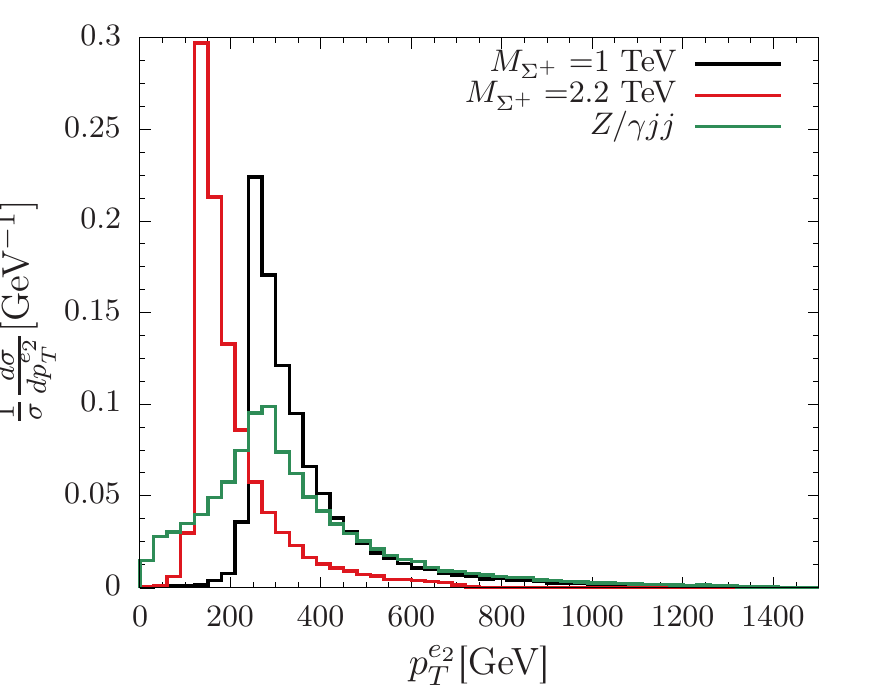}
\includegraphics[width=0.46\textwidth]{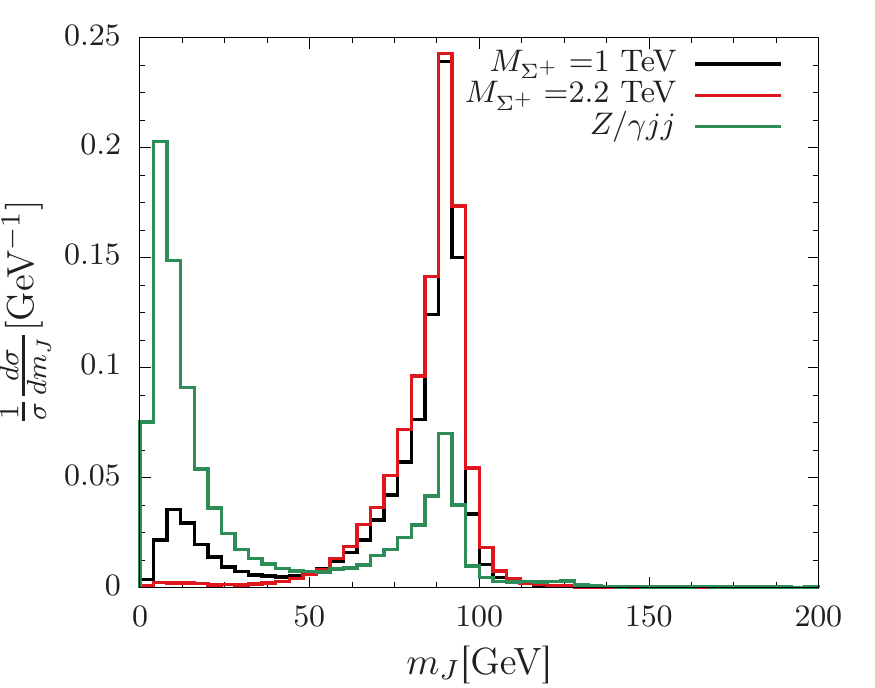}
\includegraphics[width=0.46\textwidth]{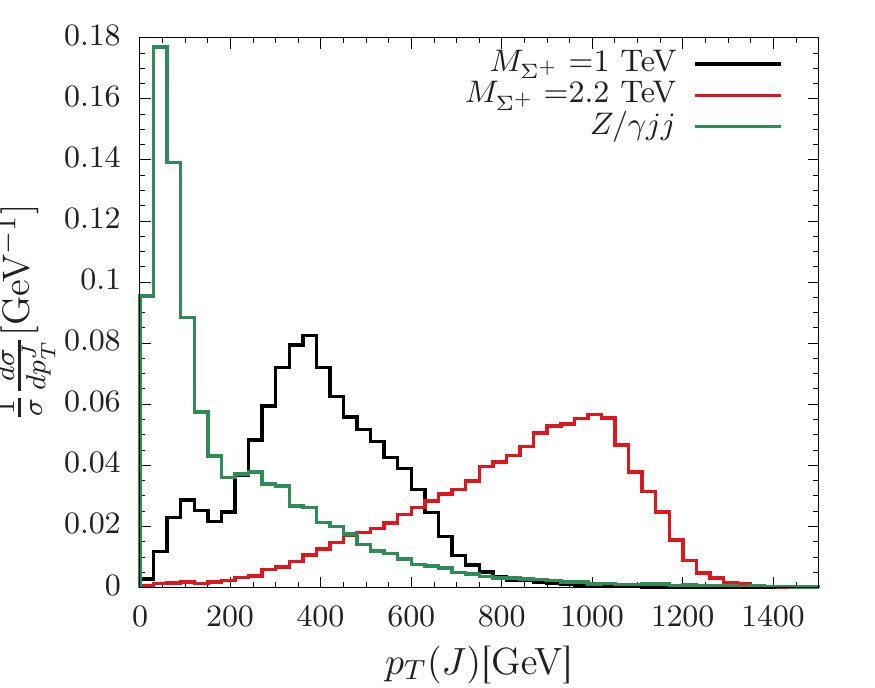}
\caption{Normalized $p_T^{e_1}$ (upper left), 
$p_T^{e_2}$ (upper right), $m_J$ (lower left), $p_T^{J}$ (lower right) distributions before applying any selection cuts 
for the  $e^- e^+ \to e^- \Sigma^{+}\to e^- e^+ Z \to e^- e^+ J $ process. The red and black distribution
corresponds to signal for $M_{\Sigma} = 1$ TeV and $M_{\Sigma} = 2.2$ TeV, while the green 
distribution is for the dominant $Z/\gamma jj$ background.}
\label{fig:TM}
\end{figure}

In this section we discuss the potential to test $\Sigma^{+}$ in 
$e^- e^+ \to e^- \Sigma^{+}$ mode (conjugate process implied) followed by $\Sigma^{+} \to e^+ Z \to e^- e^+ J$ at the $\sqrt{s}=3$ TeV $e^-e^+$ collider.
The corresponding Feynman diagram is given in Fig~\ref{ILC-Fat-3}.
There exist several SM backgrounds for this process. 
The dominant background arises from $Z/\gamma jj$ whereas,
$t\bar t$, $WWZ$, $WWjj$ and, $WZjj$ constitute subdominant contributions.
In order to find the optimized cuts we first plotted normalized distributions of the
leading, subleading electrons and the leading fat jet in Fig.~\ref{fig:TM} for signal and leading SM background
for two benchmark points $M_{\Sigma} = 1$ TeV and $2.2$ TeV respectively. 
Note that as in before these distributions are generated with basic cuts.

Our focus of interest is $800$ GeV $\leq M_{\Sigma} \leq 2.9$ TeV.
Based on the normalized distributions in Fig.~\ref{fig:TM},
we have applied following set of advanced selection cuts:
\begin{enumerate}
 \item the fat jet transverse momentum is required to be $p_T^{J} >150$ GeV.
 \item the jet mass should be $80~\mbox{GeV} < m_J < 100$ GeV.
 \item the $p_T$ of the leading electron is chosen to be $p_T^{e_1} >350$ GeV, while
$p_T$ of the subleading electron, $p_T^{e_2}$ assumed to be  $>200$ GeV for $900$ GeV $\leq M_{\Sigma} \leq 1.4$ TeV, 
$>100$ GeV for $1.6$ TeV $\leq M_{\Sigma} \leq 2.4$ TeV and, $>40$ GeV for  $2.6$ TeV $\leq M_{\Sigma} \leq 3.0$ TeV.
\end{enumerate}
%%%%%%%%%%%%%%%%%%%%%%%%%%%%%%%%%%%%
 For the chosen $M_{\Sigma}$ range the subleading recoils against the $\Sigma^{+}$.
Therefore we found that for the heavier $\Sigma^{+}$ the significances improves significantly with lower $p_T^{e_2}$. 
The impact of the event selection cuts are provided in Table~\ref{table:TM} for $M_{\Sigma} = 1$ TeV for illustration.
%In Fig~\ref{mixing from eeJ:TM} we plotted the discovery and evidence contours for three different integrated luminosities
%$\mathcal{L}=1000,3000,5000\,\text{fb}^{-1}$

\begin{table}[!htbp]
\centering
\begin{tabular}{|c| c| c |c |c | c| c | c | c | c| }
\hline
 &&&&&&&\\  
 Selection                                         & Signal  & $Z/\gamma jj$  & $t\bar t$  & $WWZ$  & $WWjj$ & $WZjj$           & Total     \\
       cut                                         &  (fb)    & (fb)           & (fb)      & (fb)   & (fb)   & (fb)             &  Background(fb) \\             
\hline
\hline
      Basic cuts                                   &  0.105    & 7.99          & 0.01       & 0.31   & 0.11  & 0.14              & 8.56 \\
      $p_T^{e_1} >350$ GeV, $p_T^{e_2} >200$ GeV   &  0.092   & 4.59          & 0.001      & 0.06   & 0.02  & 0.03              & 4.7  \\
      $p_T^{J} >150$ GeV                           &  0.085    & 1.82          & 0.0002     & 0.04   & 0.02  & 0.02              & 1.9    \\
      $80~\mbox{GeV}< m_J < 100$ GeV               &  0.057     & 0.8           & 0.00002    & 0.02   & 0.01  & 0.01              & 0.84   \\
\hline
\hline
\end{tabular}
\caption{Cut flow of the signal and backgrounds for a benchmark $M_{\Sigma} = 1$ TeV for the signal $ e^- e^+ J $ at the $e^- e^+$ collider.
process.  
%with $|V_e|=5\times10^{-2}$. The CM energy is $\sqrt{s}=3$ TeV.
}
\label{table:TM}
\end{table}
%%%%%%%%%%%%%%
\subsection{Analysis for the final state  $2J+p_T^{miss}$ at $\sqrt{s}=3$ TeV $e^{-}e^{+}$ collider}
%%%%%%%%%%%%%%%%%%%%%%%%%%%%%%%%%%%%
\begin{figure}[]
\centering
\includegraphics[width=1\textwidth]{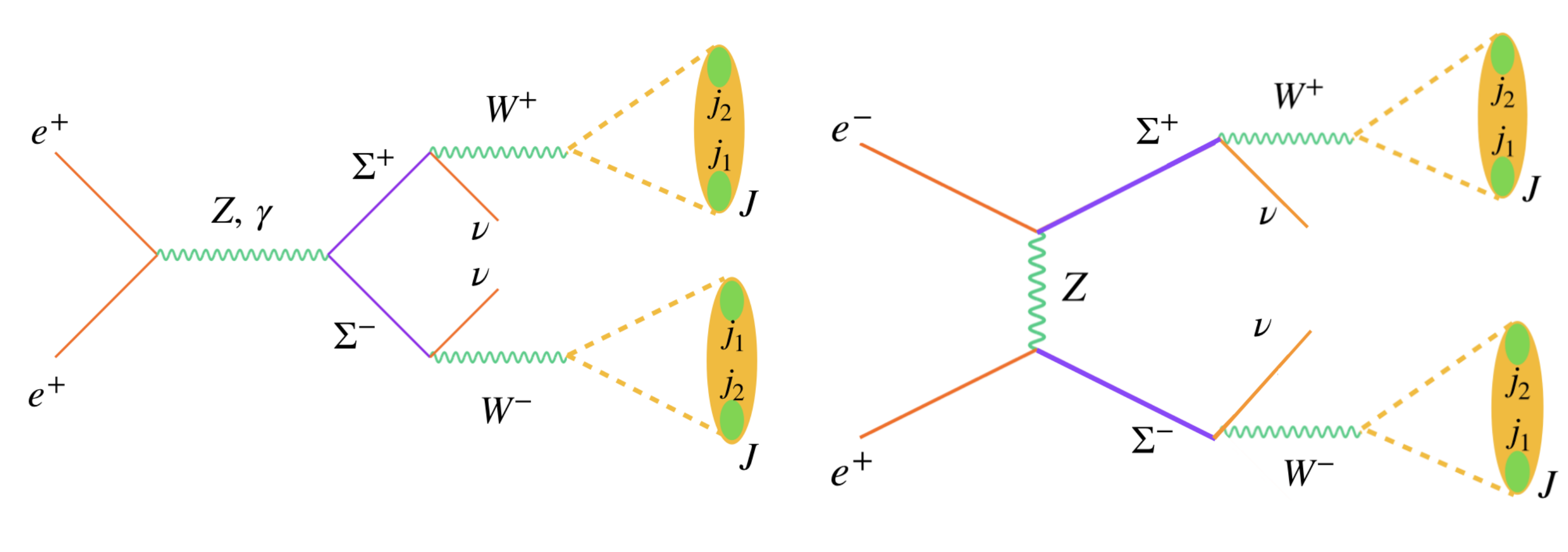}
\caption{Fat jet $(J)$ production from the $\Sigma^\pm$ at the $e^+ e^-$ collider from the $s$-channel process.}
\label{ILC-Fat-4}
\end{figure}
%%%%%%%%%%%%%%%%%%%%%%%%%%%%%
This final state arises from the pair production of $\Sigma^{+}\Sigma^{-}$ and its subsequent decay $\Sigma^{\pm}\to W^{\pm}\nu$, $W^{\pm}\to J$ at the $e^-e^+$ collider. 
The corresponding Feynman diagram is given in Fig.~\ref{ILC-Fat-4}. For this final state the dominant SM backgrounds come from the intermediate processes $WW\nu\nu$ and $WWZ$. In this case we have considered $800$ GeV $\leq M_{\Sigma} \leq 1.4$ TeV at $\sqrt{s}=3$ TeV $e^-e^+$ collider.
%%%%%%%%%%%%%%%%%%%%%%%%%%%%%%%%%%%%%%%%%%%%%%%%%%%
\begin{table*}[!htbp]
\begin{tabular}{|c|c|c|c|c|} 
\hline
Cuts & Signal[fb] & \multicolumn{2}{|c|}{Background[fb]} & Total[fb] \\ \hline
     &          & $WW\nu\nu$ & $WWZ$ &  \\ \hline
Basic Cuts  & 0.341   & 16.659 & 1.406 &  18.065 \\
 $p_{T}^{miss}>200$ GeV  &  0.303  & 9.251      &    0.941    &  10.191  \\
$p_T^{J}>300$ GeV & 0.296 & 5.813 & 0.896 & 6.708 \\
 $70\,\text{GeV}\leq m_{J}\leq 90$ GeV & 0.233 & 3.695 &  0.657   & 4.352        \\ \hline
\end{tabular}
\caption{Cut flow of the signal and background cross-sections for the final state $2J+p_T^{miss}$ for $M_{\Sigma}=1$ TeV at the $\sqrt{s}=3$ TeV $e^-e^+$ collider. 
}
\label{2J+MET 1 TeV}
\end{table*}  
%%%%%%%%%%%%%%%%%%%%%%%%%%%%%%%%%%%%%%%%%%%%
In addition to the basic cuts we have applied the following set of advanced selection cuts to reduce the SM backgrounds:
\begin{enumerate}
\item missing energy, $p_{T}^{miss}>200$ GeV.
\item transverse momentum for fat jet should be, $p_{T}^{J}>300$ GeV.
\item the fat jet mass should be, $70\,\text{GeV}\leq m_{J}\leq 90$ GeV.
\end{enumerate}
%%%%%%%%%%%%%%%%%%%%%%%%%%%

\begin{figure}[]
\centering
\includegraphics[width=0.47\textwidth]{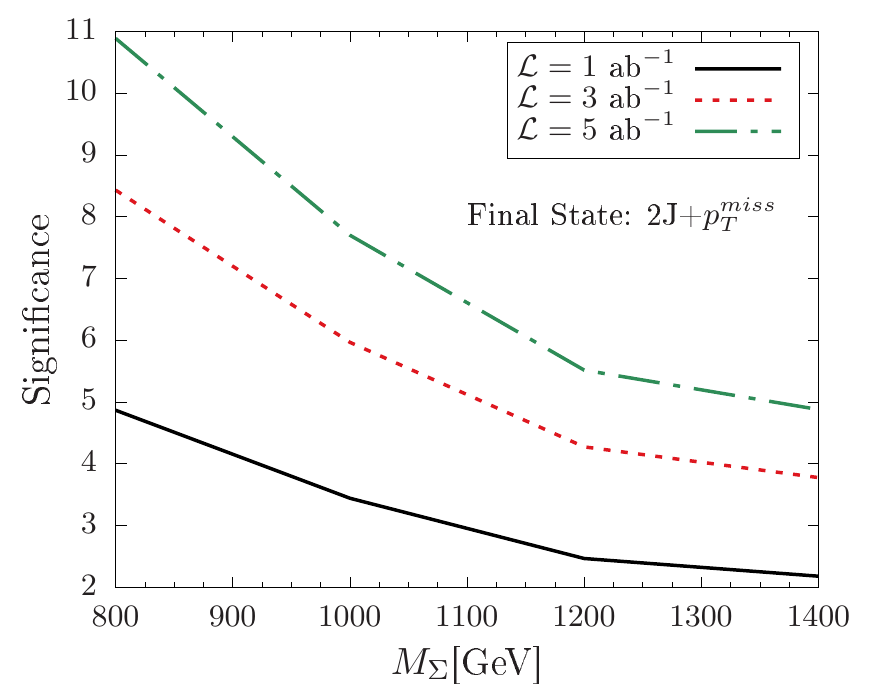}
\includegraphics[width=0.47\textwidth]{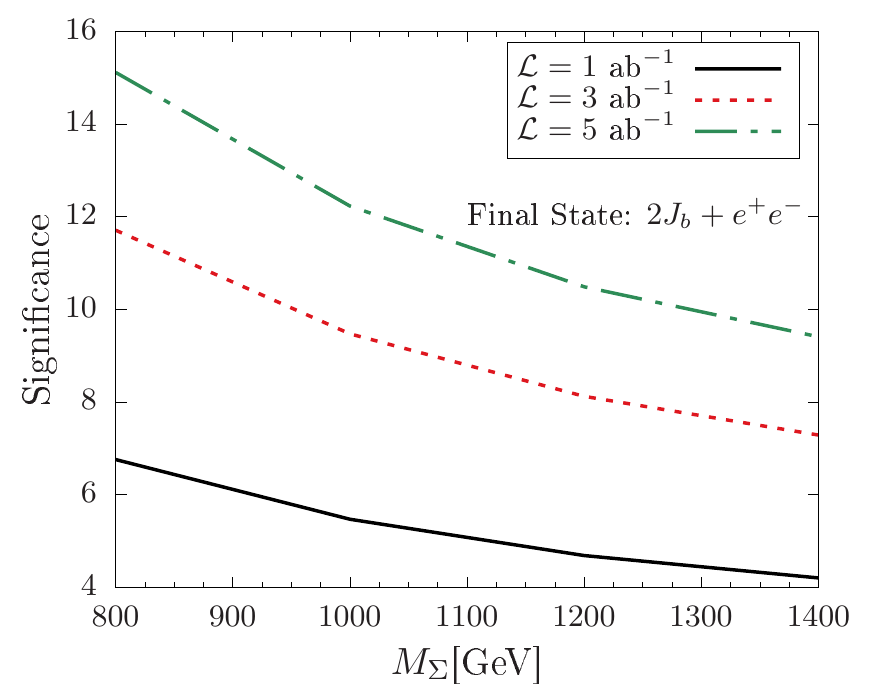}
\caption{Signal significance for the final states $2J+p_T^{miss}$~(left panel) and $2J_b+e^-e^+$~(right panel) from the pair production of $\Sigma^{\pm}$ with luminosity $1$, $3$ and $5$ $\text{ab}^{-1}$ respectively at the $\sqrt{s}=3$ TeV $e^-e^+$ collider.}
\label{significance for 2J+MET}
\end{figure}

Studying the signal and the backgrounds we have calculated the significance of the $2J+p_T^{miss}$ process and it is shown in the left panel of the Fig.~\ref{significance for 2J+MET} at the $3$ TeV $e^-e^+$ collider. The testing potential of this channel can reach up to more than $5$-$\sigma$ for $3$ ab$^{-1}$ and $5$ ab$^{-1}$ luminosities, however, the impressive significance well above $3$-$\sigma$ can reach at $1$ ab$^{-1}$ luminosity. \footnote{We use $S_\text{sig}=\frac{S}{\sqrt{S+B}}$ to calculate the signaficance.}

At this point we mention that we have also studied the $2 J_b+e^-e^+$ signal from Fig.~\ref{ILC-Fat-4} where $\Sigma^\pm$ decays into $\ell^\pm h$. Each of $h$ boson can further  decay into a pair of collimated $b$-jets. These collimated $b$ jets form a fat b-jet. Using a $70\%$ b-tagging efficiency for the signal and SM backgrounds we give only the significance of this process just as a reference in right panel of Fig.~\ref{significance for 2J+MET}. The significance of this process can prospectively reach above the $5$-$\sigma$ level at the $1$ ab$^{-1}$ luminosity at $3$ TeV $e^-e^+$ collider. Higher luminosities at $3$ ab$^{-1}$ and $5$ ab$^{-1}$ can improve the significance of the heavier triplets leading to a prospect of above $5$-$\sigma$ significance. In this context we mention that there is $J_b+e^-e^+$ final state which can be obtained from the single Higgs production as given in Fig.~\ref{ILC-Fat-2}. Due to the single fat b-jet this channel is less efficient compared to the $2 J_b+e^-e^+$ signal.
%%%%%%%%%%%%%%%%%%%%%%%%%%%
%%%%%%%%%%%%%%%%%%%%%%%%
\subsection{Analysis for the final state $e^{\pm}+J+j_1$ at FCC-he}
%%%%%%%%%%%%%%%%%%%%%%%%%%%%%%%%%%%%%%%%%%%%%%%%%%%%%
\begin{figure}[]
\centering
\includegraphics[width=0.75\textwidth]{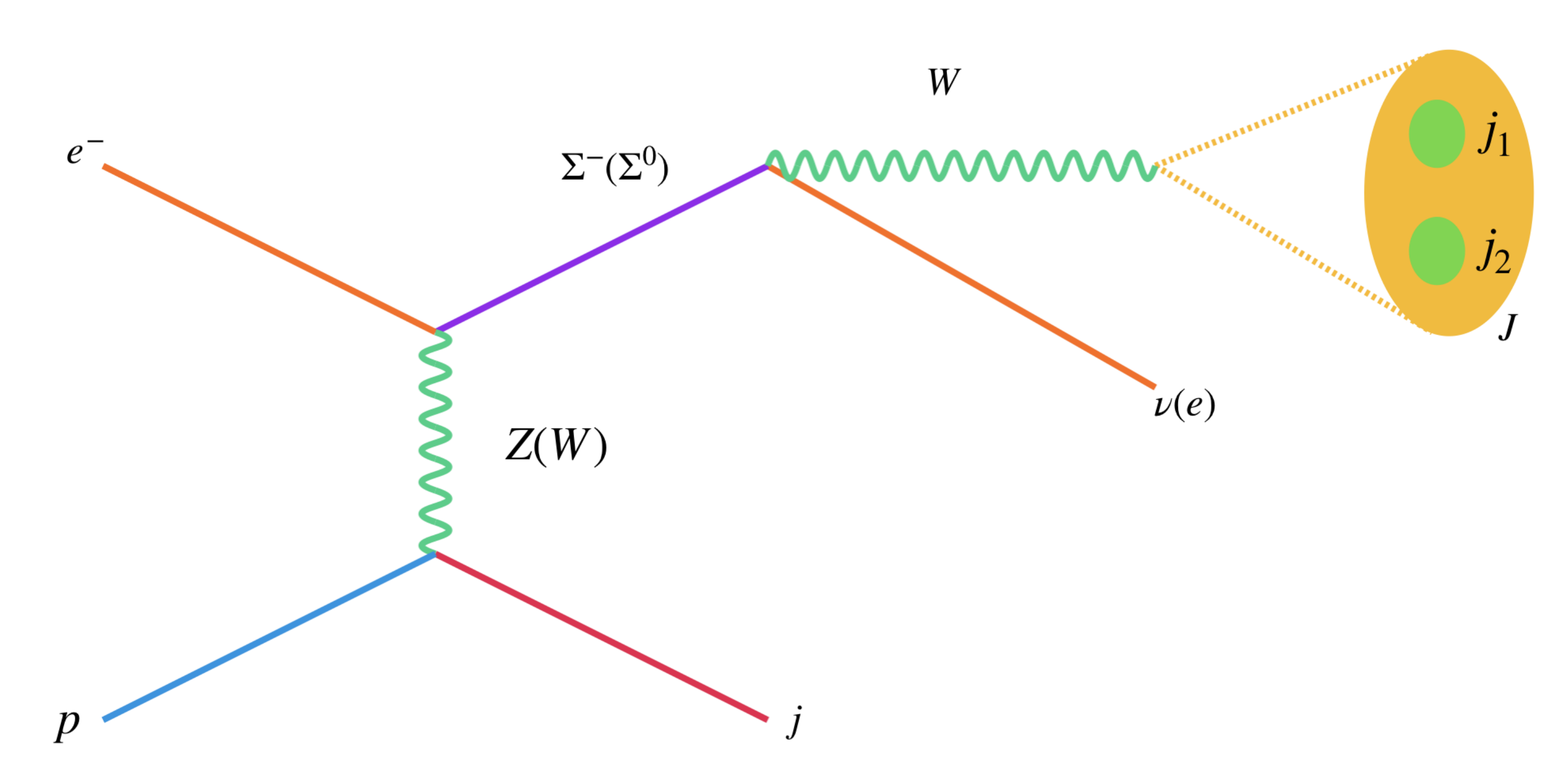}
\caption{Fat jet $(J)$ production from the $\Sigma^-$ and $\Sigma^0$ at the $e^- p$ collider from the $t$-channel process.}
\label{LHeC-Fat-1}
\end{figure}
%%%%%%%%%%%%%%%%%%%%%%%%%%%%%
This final state arises from the production of $j_1\Sigma^0$ and the subsequent decay of $\Sigma^0\to e^{\pm}W^{\mp}$, $W^{\mp}\to J$ at the $e^-p$ collider.
The corresponding Feynman diagram is given in Fig.~\ref{LHeC-Fat-1}. In this case we consider a $\sqrt{s}=3.46$ TeV $e^-p$ collider. 
In this section we study the visible particles in the final sate.
Here we have two different process $e^{+}+J+j_1$ and $e^{-}+J+j_1$. The first one is Lepton number violating (LNV) and second one is Lepton number conserving (LNC) process at the $e^-p$ collider. We have combined both these LNV and LNC processes to obtain the final state $e^{\pm}+J+j_1$. 
%%%%%%%%%%%%%%%%%%%%%%%%%%%%%%%%%%%

We expect the LNV signal to be almost background free, unless some $e^++\text{jets}$ events appear from the radiation effects, which, one expects to be negligible. 
For the LNC channel, the dominant SM backgrounds come from the SM processes $e^-jjj$, $e^-jj$ and $e^-j$ including the initial and final state radiations. 
We have shown the normalized distributions of the leading lepton $p_T$, fat jet $p_T$, invariant mass of fat jet, invariant mass of leading lepton and fat jet system in Fig.~\ref{ephist}. These distributions are for two benchmark points $M_\Sigma=1$ TeV and $1.5$ TeV. For this final state we focus in the mass range $800$ GeV $\leq M_{\Sigma} \leq 2$ TeV. 
In view of these distributions in Fig.~\ref{ephist}, we apply the following set of advanced selection cuts: 
%%%%%%%%%%%%%
\begin{figure}[]
\centering
\includegraphics[width=0.47\textwidth]{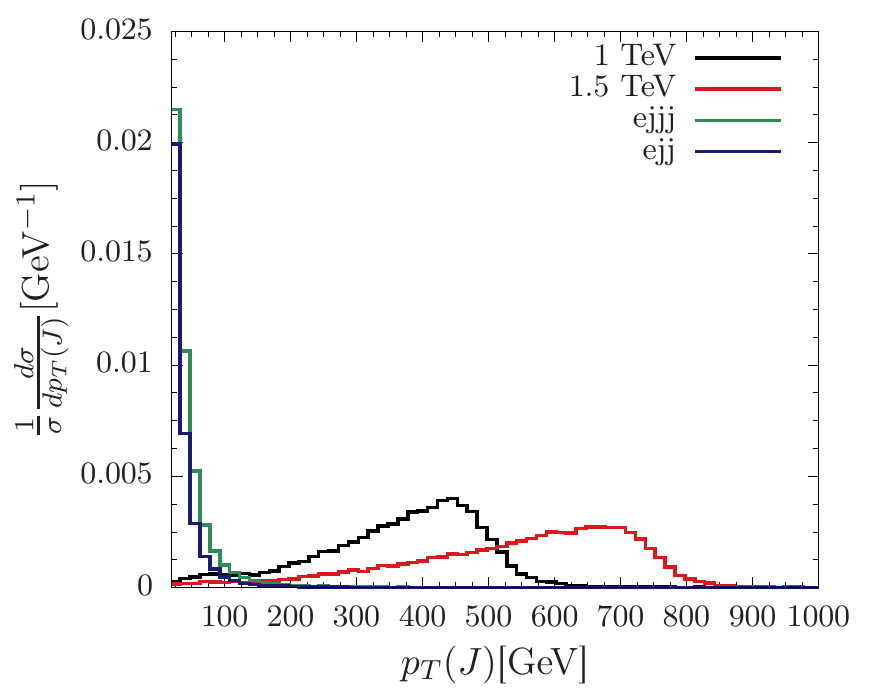}
\includegraphics[width=0.47\textwidth]{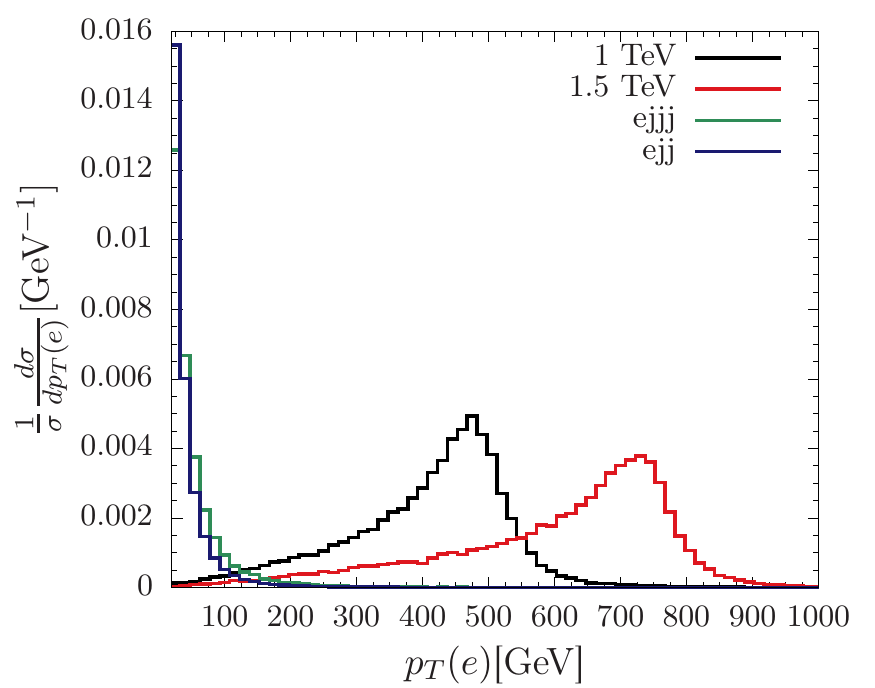}
\includegraphics[width=0.47\textwidth]{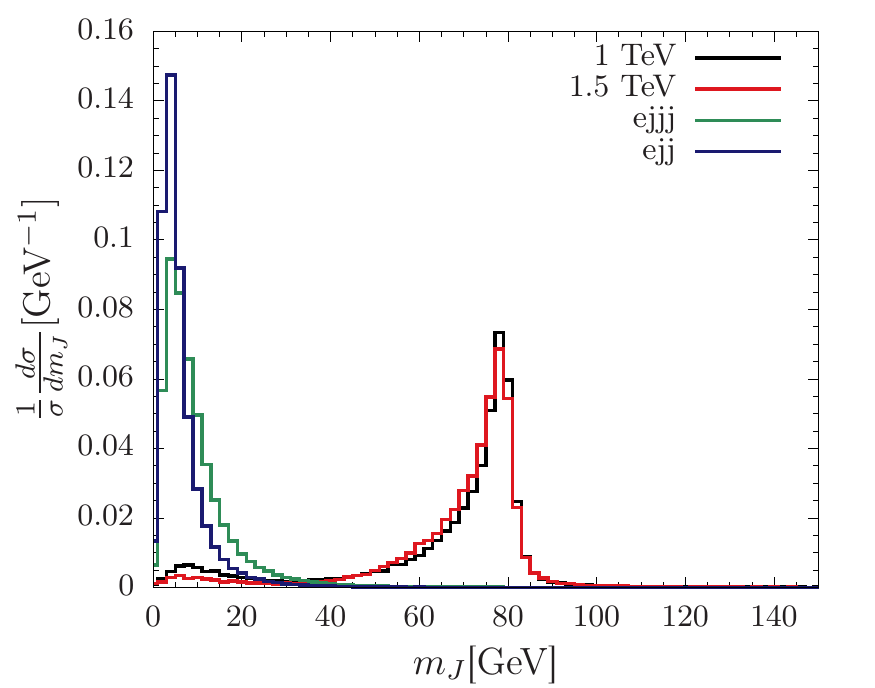}
\includegraphics[width=0.47\textwidth]{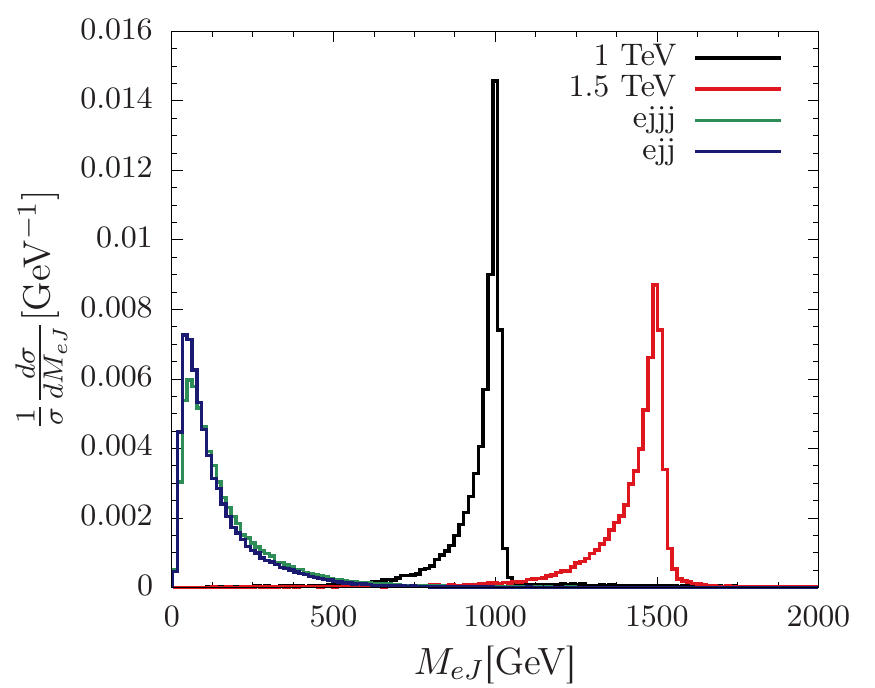}
\caption{Normalized fat jet $p_T$, leading lepton $p_T$, fat jet invariant mass and invariant mass of fat jet and lepton distributions for the final state $e^{\pm}+J+j_1$. The black and red line corresponds to signal for $M_{\Sigma}=1$ TeV and 1.5 TeV, respectively.}
\label{ephist}
\end{figure}
%%%%%%%%%%%%%%%%%%%%%%%
\begin{enumerate}
\item transverse momentum of fat jet $p_{T}^{J}>200$ GeV for $800$ GeV $\leq M_{\Sigma} \leq 1.4$ TeV and $p_T^{J}>400$ GeV for $1.5$ TeV $M_{\Sigma} \leq 2.0$ TeV.
\item transverse momentum of lepton, $p_{T}^{e^{\pm}}>200$ GeV for $800$ GeV $\leq M_{\Sigma} \leq 1.4$ TeV and $p_T^{e^{\pm}}>400$ GeV for $1.5$ TeV $\leq M_{\Sigma} \leq1.5$ TeV.
 \item fat jet mass $70\,\text{GeV}\leq m_{J}\leq 90$ GeV.
 \item invariant mass window of $(e^{\pm})$ and fat jet $(J)$ system, $|M_{eJ}-M_{\Sigma}|\leq 20$ GeV.
\end{enumerate}
%%%%%%%%%%%%%%%%
\begin{table*}[!htbp]
\begin{tabular}{|c|c|c|c|c|} 
\hline
Cuts & Signal[fb] & \multicolumn{2}{|c|}{Background[fb]} & Total[fb] \\ \hline
     &          & $ejjj$ & $ejj$ &  \\ \hline
Basic Cuts  & 0.337   & $2.905\times 10^5$ & $5.404\times 10^5$ &  $8.309\times 10^5$ \\
 $p^{J}_{T}>200$ GeV  &  0.303  & $2.592\times 10^3$      &    $1.799\times 10^3$    &  $4.391\times 10^3$  \\
$p_T^{e}>200$ GeV & 0.294 & $1.891\times 10^3$ & $1.449\times 10^3$ & $3.34\times 10^3$ \\
 $70\,\text{GeV}\leq m_{J}\leq 90$ GeV & 0.202 & 351.808 &  205.075   &  $556.883$     \\
$|M_{eJ}-M_{\Sigma}|\leq 20$ GeV & 0.134 & 9.138  &  4.661   & 13.799   \\\hline
\end{tabular}
\caption{Cut flow of the signal and background cross-sections for the final state $e^{\pm}+J+j_{1}$ for $M_{\Sigma}=1$ TeV with $\sqrt{s}=3.46$ TeV $e^-p$ collider.
% for the mixing $V_e=0.05$.
}
\label{ep collider 1 TeV}
\end{table*}  
%%%%%%%%%%%%%%%%%
Note that, $70\,\text{GeV}\leq m_J\leq 90\,\text{GeV}$ cuts out the SM backgrounds low energy peaks. Similarly high $p_T$ cuts for leading lepton and fat jet are extremely useful to reduce the SM backgrounds.

It is difficult to obtain fat jet for the background process $ej$ because of the $t$ channel exchange of the $Z$ boson and photon. Initial and final state radiations can give low energy jets which can produce soft fat jet. Therefore $ej$ background can completely be reduced with high $p_T^e$, $p_T^J$ cuts and fat jet mass $70\,\text{GeV}\leq m_J\leq 90\,\text{GeV}$. These cuts will not be enough to reduce the irreducible backgrounds coming from the process $ejj$ and $ejjj$. However, both of these backgrounds can be reduced using the invariant mass cut of the $\Sigma^0$. As $\Sigma^0$ decays according to $\Sigma^0\to e W, W \to J$, the invariant mass of  $eJ$ system should be close to mass of the $\Sigma^0$. 
Therefore an invariant mass window cut $|M_{eJ}-M_\Sigma|\leq 20$ GeV will be extremely useful to reduce these two SM backgrounds further.
%%%%%%%%%%%%%%%%%%%%%%%%%%%%%%%%
\begin{table*}[!htbp]
\begin{tabular}{|c|c|c|c|c|} 
\hline
Cuts & Signal[fb] & \multicolumn{2}{|c|}{Background[fb]} & Total[fb] \\ \hline
     &          & $ejjj$ & $ejj$ &  \\ \hline
Basic Cuts  & 0.060   & $2.905\times 10^5$ & $5.404\times 10^5$ &  $8.309\times 10^5$ \\
 $p^{J}_{T}>400$ GeV  &  0.048  & 274.136      &    167.788    &  441.924  \\
$p_T^{e}>400$ GeV & 0.047 & 203.318 & 135.163 & 338.481 \\
 $70\,\text{GeV}\leq m_{J}\leq 90$ GeV & 0.033 & 22.844 &  18.643   &  41.488     \\
$|M_{eJ}-M_{\Sigma}|\leq 20$ GeV & 0.015 & 0.115  &  0.023   & 0.138   \\\hline
\end{tabular}
\caption{Cut flow of the signal and background cross-sections for the final state $e^{\pm}+J+j_{1}$ for $M_{\Sigma}=1.5$ TeV with $\sqrt{s}=3.46$ TeV $e^-p$ collider.
% for the mixing $V_e=0.05$.
}
\label{ep collider 1.5 TeV}
\end{table*}  
%%%%%%%%%%%%%%%%%%%%%%%%%%%%%%%%%%%%
We have shown the cut flow for two benchmark points $M_{\Sigma}=1$ TeV and 1.5 TeV in Table.~\ref{ep collider 1 TeV} and \ref{ep collider 1.5 TeV}.

Further we also comment that $\Sigma^-j$ channel is show a signature of $j_1+J+p_T^{miss}$ which is less significant compared to the $j_1+e^\pm+J$ channel due to the absence of the visible lepton.
The presence of the jets and the missing energy will have more SM backgrounds which will make the $j_1+J+p_T^{miss}$ final state less sensitive.
%%%%%%%%%%%%%%%%%%%%%%%%%%%%%%%%%%%%%%
\section{Discussions}
\label{Dis}
\begin{figure}[]
\centering
\includegraphics[width=0.495\textwidth]{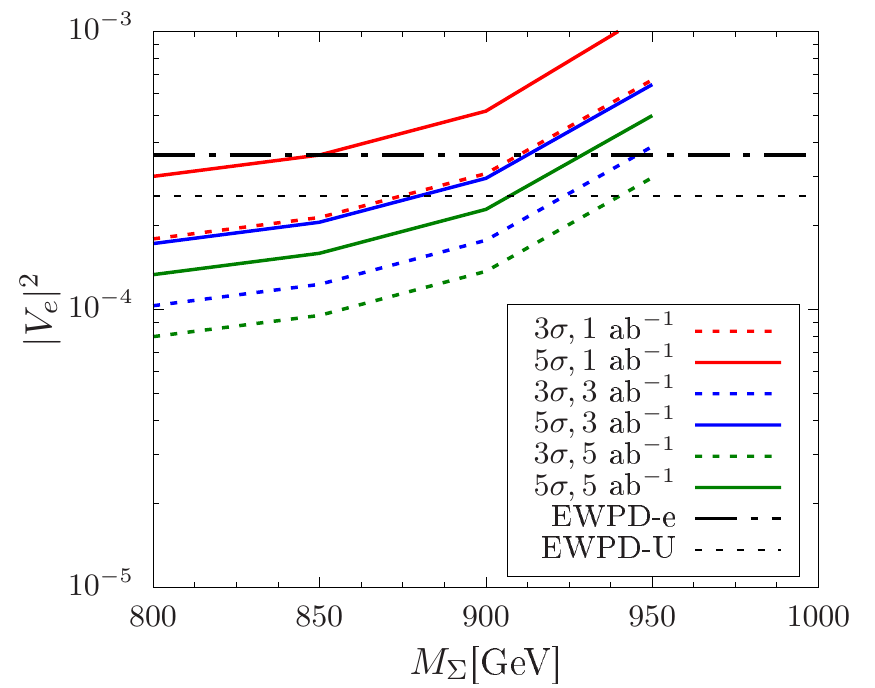}
\includegraphics[width=0.495\textwidth]{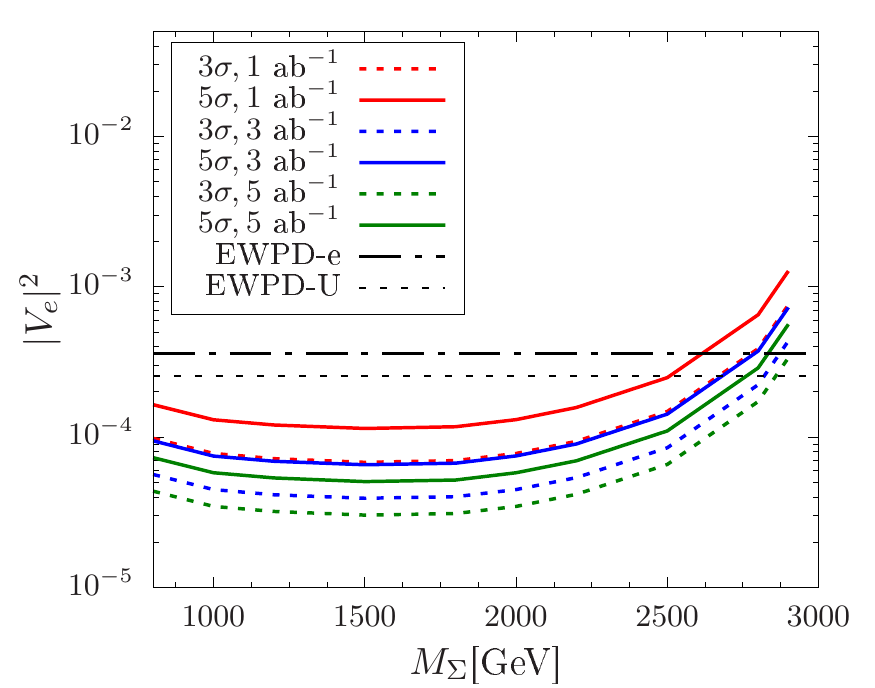}\\
\includegraphics[width=0.495\textwidth]{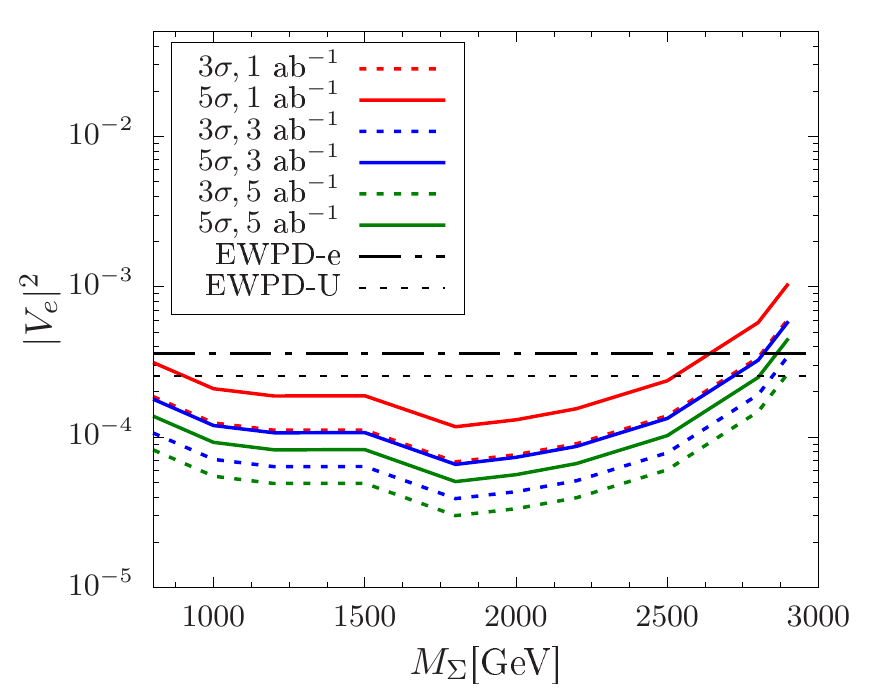}
\includegraphics[width=0.495\textwidth]{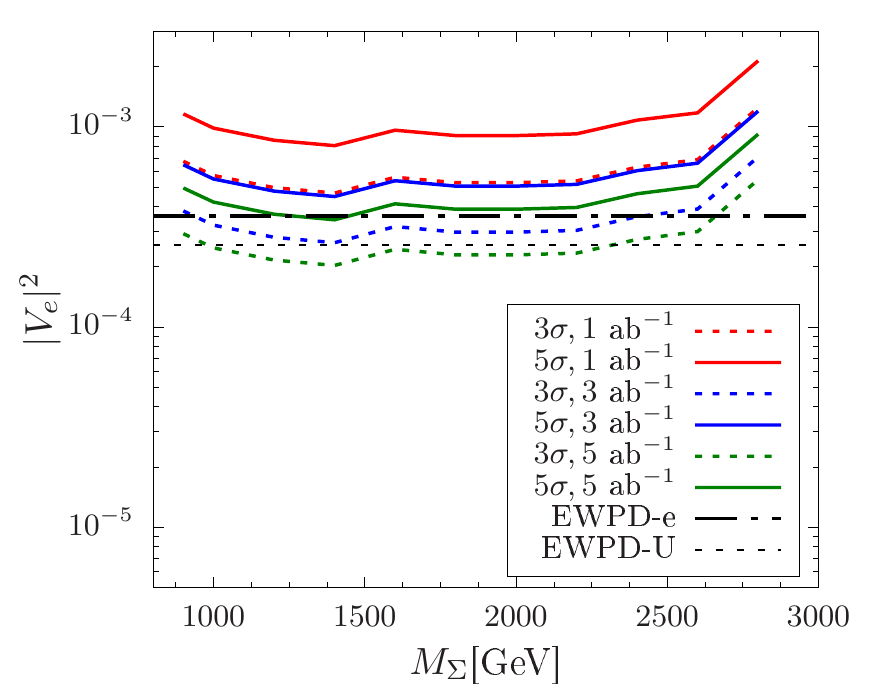}
\caption{Sensitivity reach of the mixing angle with the luminosities $\mathcal{L}=1$ ab$^{-1}$, 3 ab$^{-1}$ and $5$ ab$^{-1}$ at $3\sigma$ and $5\sigma$ significance from the final state $e^{\pm}+J+p_{T}^{miss}$ at $\sqrt{s}=1$ TeV (top, left) and $\sqrt{s}=3$ TeV (top, right), respectively at the $e^-e^+$ collider. The same for the final states $J_{b}+p_{T}^{\rm miss}$ (bottom, left) and $e^{+}+e^{-}+J$ (bottom, right), respectively at the $\sqrt{s}=3$ TeV $e^-e^+$ collider with same luminosities like the top panel. The limit from the EWPD for the electron flavor has been represented by the black dot-dashed line from \cite{delAguila:2008pw}.}
\label{ILCmix}
\end{figure}

\begin{figure}[]
\centering
\includegraphics[width=0.8\textwidth]{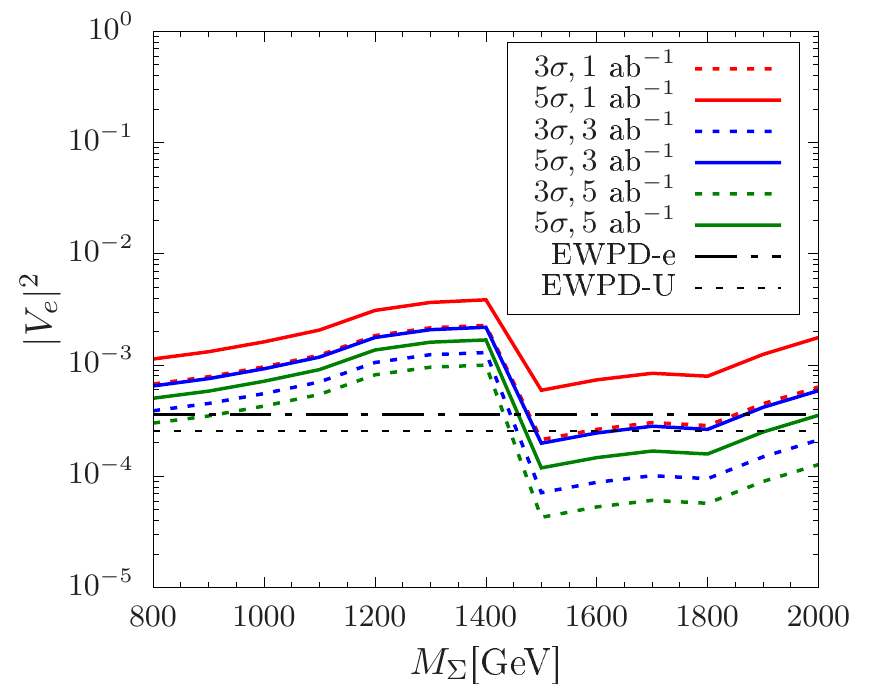}
\caption{Sensitivity reach of the mixing angle with the luminosities $\mathcal{L}=1$ ab$^{-1}$, 3 ab$^{-1}$ and $5$ ab$^{-1}$ at $3\sigma$ and $5\sigma$ significance respectively from the final state $e^{\pm}+J+j_1$ at $\sqrt{s}=3.46$ TeV of  the $e^-p$ collider. 
}
\label{ep collide mixing}
\end{figure}

%%%%%%%%%%%%%%%%%%%%%%
After studying the signals and the SM backgrounds for the triplet fermion production at the $e^-e^+$ collider and $e^-p$ at different energies and luminosities we calculate the bounds on the mass mixing plane at $3$-$\sigma$ and $5$-$\sigma$ significance. To calculate the bounds on the mixings from the $e^-e^+$ collider we use two different center of mass energies like $1$ TeV and $3$ TeV. To do the same at the $e^-p$ collider we use $3.46$ TeV center of mass energy. At these colliders we have used $1$ ab$^{-1}$, $3$ ab$^{-1}$ and $5$ ab$^{-1}$ luminosities respectively. We compare our results with the bounds obtained from \cite{delAguila:2008pw}. We use the bounds on the mixing $V_e=0.019$ and the universal bounds $0.016$ as studied from the Electroweak Precision Data (EWPD). We represent the bounds as EWPD-e and EWPD-U in Figs.~\ref{ILCmix} and \ref{ep collide mixing} by the horizontal dot-dashed and dotted lines respectively. 

%%%%%%%%%%%%%%%%%%%%%%%%%%%%%
%%%%%%%%%%%%%
The bounds obtained from the $e^-e^+$ collider studying a variety of final states are shown in Fig.~\ref{ILCmix}.
We have studied the $e^\pm+J+p_T^{miss}$ final state from Fig.~\ref{ILC-Fat-1} at the $1$ TeV (top, left)  and $3$ TeV (top, right) $e^-e^+$ colliders.
One can probe the mixing up to $8\times 10^{-5}$ at the $1$ TeV collider at $3$-$\sigma$ significance with the luminosity of $5$ ab$^{-1}$.
With the same luminosity, the bounds remain below the EWPD-e and EWPD-U up to $900$ GeV (EWPD-U) and $945$ GeV (EWPD-e) at $5$-$\sigma$ significance.
The same has been studied at the $3$ TeV $e^-e^+$ collider where the results can impressively be improved even with $1$ ab$^{-1}$ luminosity with $5$-$\sigma$ significance up to $2.5$ TeV triplets which can be further improved at the higher luminosity probing the heavy triplets up to $2.9$ TeV. We have studied the $J_b+p_T^{miss}$ final state from Fig.~\ref{ILC-Fat-2} and the corresponding bounds at the $3$ TeV $e^-e^+$ collider are given in Fig.~\ref{ILCmix} (bottom, left). The lowest mixing $3.5\times 10^{-5}$ could be reached $1.725$ TeV heavy triplet mass. 
%Heavier mass can boost the Higgs well which helps in creating fat b-jets. 
At the collider energy threshold the cross section decreases at the $e^-e^+$ collider which in turn does not help the further heavier triplets in getting lower bounds, however, analyzing the signal and the corresponding SM backgrounds we find that sightly heavier triplets than $2.5$ TeV can be probed with $1$ ab$^{-1}$ luminosity with $5$-$\sigma$ significance. Higher luminosity will make the results better. We have studied $e^+ e^-+J$ final state from Fig.~\ref{ILC-Fat-3} and the bound obtained from this final state have been shown in Fig.~\ref{ILCmix} (bottom, right) where the best results can be obtained at the $3$ TeV $e^-e^+$ collider with $5$ ab$^{-1}$ luminosity just below the limits obtained from the EWPD at $3$-$\sigma$ significance.

Bounds obtained from the $e^-p$ collider have been shown in Fig.~\ref{ep collide mixing} studying the $e^\pm+J+j_1$ final state from Fig.~\ref{LHeC-Fat-1} which has a visible particle in the final state from the leading decay mode compared to the invisible one from the other possibilities. In this case we find that the heavy triplets are favored to study the fat jets due to 
%%%%%%%%%%%%%%%%%%%%%%%%%%%%%%%%%%%%%%%%%%
%%%%%%%%%%%%%%%%%%%%%%%%%%%%
their better efficiency for boosting the $W$ boson coming from the leading decay mode of the $\Sigma^0$. Comparing our results with the bounds obtained from the EWPD we find that   
higher luminosity with $3$ ab$^{-1}$ and $5$ ab$^{-1}$ will be useful to probe heavier triplets. Triplets heavier than $1.4$ TeV can be probed up to the mixings $3\times 10^{-5}$ at $3$-$\sigma$ significance and $10^{-4}$ at the $5$-$\sigma$ significance at $5$ ab$^{-1}$ luminosity respectively. 
%In both of the cases the triplet mass is $1.5$ TeV. 
%The bounds obtained from the heavier mass are below the limits from the EWPD but less strong compared to the lighter side between $1.5$ TeV to $1.9$ TeV. This is because of the cross section of the triplet production at the $e^-p$ collider which reduces with $M_\Sigma$. 
The bounds obtained for $M_\Sigma > 1.4$ TeV is better than those obtained for $M_\Sigma <1.4$ TeV. This is because of the heavier triplets which produced fat jets better than the comparatively lighter ones.  

%%%%%%%%%%%%%%%
\section{Conclusion}
\label{conclusion}
%%%%%%%%%%%%%
We have studied the triplet fermion initiated seesaw model which is commonly known as the type-III seesaw scenario which is responsible for the light neutrino mass generation through the seesaw mechanism. As a result a light-heavy mixing appears in the model. We consider the production of the charged and neutral multiplets of the triplet fermion at the $e^-e^+$ and $e^-p$ colliders at different center of mass energies. Being produced in association with the SM particles and as well as in pair the charged and neutral multiplets can decay into a variety of final states. 

In this article we consider mostly the leading decay modes of the triplets and some next-to-leading modes, too. As the mass of the triplet is a free parameter, we consider the heavier mass of the triplet which can sufficiently boost its decay products. As a results the decay products including SM gauge bosons and Higgs boson can manifest fat jet through their leading hadronic decay modes. Finally  we study a variety of final states and the SM backgrounds to probe the light-heavy mixing as a function of the triplet mass for different center of mass energies at the above mentioned colliders using different integrated luminosities. Comparing our results with the bounds on the light-heavy mixing obtained from the electroweak precision test results we find that the heavier triplets can be successfully probed and the bounds on the light-heavy mixing as a function of the triplet mass can be better than the results from the electroweak test. 
%At this point we want to mention that at the $e^-p$ collider one can successfully probe the Majorana nature of the neutral multiplet. 

Apart from the associate production of the triplet we study the pair production of the heavy triplets at the $e^- e^+$ collider followed by the decays into the SM bosons.
Due to the heavier mass, the triplets can easily boost the daughter particles so that the hadronic decay products of the SM gauge bosons among them can manifest fat jet signatures 
which can efficiently segregate the signals from the SM backgrounds leading to the test above $5$-$\sigma$ significance in the near future.
%%%%%%%%%%%%%
\begin{acknowledgments}
The work of A. D. is supported by the JSPS, Grant-in-Aid for Scientific Research, No. 18F18321.
The work of S. M. is supported by the Spanish grants FPA2017-85216-P (AEI/FEDER, UE), PROMETEO/2018/165 (Generalitat Valenciana) and the Spanish Red Consolider MultiDark FPA2017-90566-REDC. The work of T. M. is supported by the grant MOST 106-2112-M-002-015-MY3.
\end{acknowledgments}
%%%%%%%%%%%%%%%%%%%%%%%%%%%%%%%%%%%%%%%%%%%
\bibliography{bibliography}
\bibliographystyle{utphys}
\end{document}